\documentclass[submission, LectureNotes]{SciPost}
\usepackage{amsmath}
\usepackage{amsfonts}
\usepackage{amssymb}
\usepackage{bm}

\begin{document}

\begin{center}{\Large \textbf{
Introduction to quantum non-reciprocal interactions: from non-Hermitian Hamiltonians to quantum master equations and quantum feedforward schemes
}}\end{center}

\begin{center}
A. A. Clerk
\end{center}

\begin{center}
Pritzker School of Molecular Engineering, University of Chicago, Chicago, USA
\\
aaclerk@uchicago.edu
\end{center}

\begin{center}
\today
\end{center}


\section*{Abstract}
{\bf
These lecture notes from the 2019 Les Houches Summer School on Quantum Information Machines provides a pedagogical introduction
to the theory of non-reciprocal quantum interactions and devices.  The goal is connect various approaches and concepts, including 
Hamiltonians encoding synthetic gauge fields,  scattering descriptions, quantum master equations, and non-Hermitian Hamiltonians.
The importance of having both non-trivial synthetic gauge fields and dissipation for obtaining non-reciprocal interactions is stressed.  Connections to broader topics such as quantum reservoir engineering and the quantum theory of continuous-measurement based feedforward are also discussed.  
}

\vspace{10pt}
\noindent\rule{\textwidth}{1pt}
\tableofcontents\thispagestyle{fancy}
\noindent\rule{\textwidth}{1pt}
\vspace{10pt}

\section{Introduction}
\label{sec:intro}

Devices that scatter incident waves (electromagnetic or acoustic) in a fundamentally asymmetric manner play a crucial role in a variety of classical and quantum applications.  Perhaps the most common devices are isolators (two-port devices which permit transmission between the ports in only one direction) and circulators (multi-port devices where transmission can only occur e.g.~from port $j$ to $j+1$, but not in the reverse direction).  Such devices are termed ``non-reciprocal", and are usually discussed in the context of the Lorentz reciprocity theorem in optics, and the Rayleigh reciprocity theorem in acoustics.  There has been in recent years a concerted effort to devise methods for achieving these kinds of non-reciprocal scattering devices without using magnetic material or magnetic fields, but instead using external driving (i.e. time modulation).  Several excellent reviews exist discussing these approaches in classical systems (see e.g.~\cite{Sounas2017,FanTutorial2020}). 

Parallel to this effort, there is growing theoretical interest in understanding the unique properties of systems whose internal dynamics are governed by effective non-Hermitian Hamiltonians which encode non-reciprocal \emph{interactions}.  Typical examples include non-Hermitian lattice models, where there is an asymmetry in, e.g., the amplitude for hopping from left to right, versus right to left \cite{Hatano1997}.  Such systems exhibit a number of unusual properties, such as the non-Hermitian skin effect, where changing boundary conditions from periodic to open completely changes the spectrum of the Hamiltonian, and localizes all eigenvectors \cite{FoaTorres2018,Yao2018,McDonald2018}.  They can also exhibit unique kinds of topological band structures \cite{Ueda2019,Ashvin2019} and can even give rise to novel phase transition physics \cite{Fruchart2021}.  The majority of work in this area assumes the existence of directional interactions as a starting point for formulating a model, without worrying about microscopic mechanisms.  In the quantum regime, this can be problematic, as it often amounts to an incomplete description of an open quantum (where one is including generalized damping effects, without accounting for the corresponding quantum fluctuations that must accompany it) \cite{McDonald2021} .

In these notes, we provide a (hopefully) pedagogical introduction for how one can microscopically achieve non-reciprocal interactions using external driving, in a way that is fully consistent quantum mechanically.  Using an extremely simple model of a three-site bosonic ring, we show explicitly how non-reciprocal scattering (as needed for an isolator or circulator) can be directly tied to non-reciprocal propagation within the ring, as described by an effective non-Hermitian Hamiltonian.  We do this in a manner which includes all relevant quantum noise effects.  This simple example highlights a general principle:  achieving non-reciprocal propagation of interactions requires both the breaking of time-reversal symmetry (in that there are non-trivial synthetic gauge fields), and requires dissipation. 

We then use this toy model to derive a quantum master equation that encodes non-reciprocal tunnelling within the ring.  This shows explicitly how non-reciprocity emerges by balancing coherent Hamiltonian interactions against the corresponding kind of dissipative interaction (as mediated by a dissipative reservoir that couples to system degrees of freedom non-locally).   
 With this example in hand, we show that the basic structure of this quantum master equation can be used to make {\it any} starting Hamiltonian interaction between two systems fully non-reciprocal.  We draw connections to both the theory of cascaded quantum sytems (where non-reciprocal interactions are generated by coupling to an external unidirectional waveguide which is then integrated out), and to quantum descriptions of measurement plus feedforward protocols (which are inherently non-reciprocal because of the one-way flow of information).  Our work thus provides a pedagogical introduction to the basic recipe for generating non-reciprocal quantum interactions introduced in Refs.~\cite{Metelmann2015} and \cite{Metelmann2017}.  It complements the analysis there in several ways (e.g.~by discussing concrete connections to non-Hermitian Hamiltonians, and by commenting on the ability of non-Hermitian interactions to generate entanglement).  

%

\section{Synthetic gauge fields}


In this section, we will introduce the first essential ingredient needed to realize non-reciprocal interactions: a synthetic gauge field, which picks out a particular direction or sense of circulation.  We will show how these can arrive by appropriate forms of driving or temporal modulation.  

\subsection{Tight binding model of coupled cavities}

Consider a collection of coupled photonic resonators (or modes), with each site $j$ having a canonical bosonic annihilation operator $\hat{a}_j$ which annihilates a photon in mode $j$; these obey the usual canonical commutation relations e.g.~$[ \hat{a}_j, \hat{a}^\dagger_k] = \delta_{jk}$, $[ \hat{a}_j, \hat{a}_k] = 0$.  The coupling between these modes is described by a photon-number conserving tight-binding (or beam-splitter) Hamiltonian having the form
\begin{equation}
	\hat{H} = \sum_j \omega_j \hat{a}^\dagger_j \hat{a}_j  
		-  \sum_{j > j'} \left( t_{jj'} \hat{a}^\dagger_j \hat{a}_{j'} + h.c. \right)
		\label{eq:HTB}
\end{equation}
Here $\omega_j$ are the resonant frequencies of each mode (i.e.~on-site energies), and the amplitudes of tunnelling between different modes is encoded in the hopping matrix elements $t_{jj'}$.  For convenience, for $j < j'$ we define $t_{jj'} = (t_{j'j})^*$

The first ingredient we will need in our minimal description is a ``synthetic gauge field".  In the context of our simple lattice model, this reduces to something simple:  we want the hopping matrix elements to have non-zero phases (i.e.~$t_{jj'} \neq t_{jj'}^* $), and we want these phases to be non-trivial, in the sense that we cannot make a gauge transformation to remove them.  More specifically, consider a local gauge transformation that shifts the phase of the annihilation operator for site $j$ by $ \theta_j$:
\begin{equation}
	\hat{a}_j \rightarrow \hat{a}_j e^{i \theta_j}
\end{equation}
The result is that after the transformation, the new hopping matrix elements become $\tilde{t}_{jj'} = t_{jj'} e^{i (\theta_j' - \theta_{j})}$.  We are interested in a situation where {\it there is no such transformation} that makes all the hopping matrix elements purely real.  
This is equivalent to the condition that there exists non-trivial effective Aharonov-Bohm phases associated with hopping around a closed loop.  For example, consider a hopping process where one hops between $1 \rightarrow 2 \rightarrow 3 \rightarrow 4 \rightarrow 1$.  
This process would be associated with an amplitude $A$ involving the product of the relevant hopping matrix elements, i.e.~
\begin{equation}
	A = t_{14} \cdot t_{43} \cdot t_{32} \cdot t_{21}
\end{equation}
If $A \neq A^*$, then we can associate an effective Aharonov-Bohm phase with this loop.  Such a phase is necessarily invariant under the local gauge transformation defined above.  The existence of phases that cannot be gauged away (leading to Aharonov-Bohm phases that are complex) can be rigorously tied to a notion of broken time reversal symmetry.  For a clear pedagogical discussion of this, we recommend Appendix A of Ref.~\cite{KochPRA2010}.

\subsection{Connection to a continuum description of a particle in a magnetic field}

Recall that such gauge invariant phases would emerge naturally if our Hamiltonian described charged particles hopping on a lattice in the presence of a magnetic field.  In this case the phases of $t_{jj'}$ can encode a true Aharanov-Bohm phase, and are determined by the electromagnetic vector potential $\vec{A}$ via the Peierls substitution:
\begin{equation}
	t_{jj'} =   |t_{jj'}| \exp \left( i \frac{q}{\hbar} \int_{\vec{R}_j'}^{\vec{R}_j} \vec{A} \cdot d \vec{s} \right)
	\label{eq:Peierls}
\end{equation}
Here $\vec{R}_j$ denotes the real space position of the $j$th site in the lattice. 
It is instructive to verify that the above lattice Hamiltonian is indeed a discretized version of the usual Hamiltonian describing a charged particle in a magnetic field, i.e.
\begin{equation}
	\hat{H}_{q} = \frac{1}{2m} \left( \hat{\vec{p}} - \frac{q}{c} \vec{A} \right)^2
	\label{eq:HMinimal}
\end{equation}
where $m$ is the mass of the particle, $c$ the speed of light, and $\vec{p}$ the momentum.  For simplicity, we focus on particles hopping on a 1D tight binding lattice with lattice constant $a$.  
Using $|j \rangle$ to denote a position eigenket centered on the lattice vector $\vec{R}_j = (ja,0,0)$, and switching to a first quantized representation for convenience, the 1D tight binding Hamiltonian with the Peierls phase has the form:
\begin{equation}
	\hat{H}_{1D} = \sum_j \left(
		-t e^{i \phi_j}  | j+1 \rangle \langle j | + h.c. \right)
		\label{eq:H1D}
\end{equation}
where from Eq.(\ref{eq:Peierls}), the the phase $\phi_j$ is given by:
\begin{equation}
	\phi_j \simeq a \frac{q}{\hbar} A_x(\vec{R}_j)
\end{equation}
We have assumed here that the vector potential $A_x(\vec{R})$ does not change significantly over a single lattice constant.  

Next, recall that real space translations are generated by the momentum operator:
\begin{equation}
	| j+1 \rangle = 
		\exp \left( - \frac{i}{\hbar} \hat{p}_x a \right) |j \rangle
\end{equation}
Let's use this to re-express the rightwards-hopping term in the Hamiltonian of Eq.~(\ref{eq:H1D}) in the limit of a small lattice constant $a \rightarrow 0$:
\begin{equation}
	e^{i \phi_j}  | j+1 \rangle \langle j |  \simeq
		\left(1 + i \phi_j - \phi_j^2 / 2 \right)
		\left(1 - \frac{i \hat{p}_x a}{\hbar} - \frac{(\hat{p}_x a)^2}{2\hbar^2} \right) |j \rangle \langle j | 
	\label{eq:Displacement}
\end{equation}
Let's define the operator $\hat{\phi} = \sum_j \phi_j |j \rangle \langle j |$.
If we now add Eq.~(\ref{eq:Displacement}) with its Hermitian conjugate, and then sum over all sites $j$, we obtain:
\begin{align}
	\sum_j \left( e^{i \phi_j}  | j+1 \rangle \langle j | + h.c. \right)  &  \simeq
	\sum_j 
		\left(1 - \frac{i \hat{p}_x a}{\hbar} - \frac{(\hat{p}_x a)^2}{2\hbar^2} \right)
		\left(1 + i \hat{\phi} - \hat{\phi}^2 / 2 \right)
		 |j \rangle \langle j | 
		+  h.c. \\
	& = 
	\left( 2 +  \{ \hat{\phi} , \frac{ \hat{p}_x a}{\hbar} \} - \hat{\phi}^2 - \left( \frac{ \hat{p}_x a}{\hbar} \right)^2 \right)   
	\\
	& = 
	\left( 2 + \frac{a^2}{\hbar^2} \left \{  q \hat{A}_x , \hat{p}_x \right \} - \frac{a^2}{\hbar^2} (q \hat{A}_x)^2 - \frac{a^2}{\hbar^2} \hat{p}_x^2 \right)
	 \\
	& = 
		\left( 2 - \frac{a^2}{\hbar^2} \left( \hat{p}_x - q \hat{A}_x \right)^2 \right) 
\end{align}
where $\hat{A}_x \equiv A_x( \hat{\vec{R}} )$ is the operator describing our vector potential.    

Our 1D tight binding Hamiltonian thus takes the form
\begin{align}
	\hat{H}_{1D} & = 
		\sum_j \left(  
			(-2t) + \frac{t a^2}{\hbar^2} 
				\left( \hat{p}_x - q \hat{A}_x \right)^2 
			\right)  
			| j \rangle \langle j | \\
		 & = 
		 	  \frac{1}{2m^*}   \left( \hat{p}_x - q \hat{A}_x \right)^2 + (\textrm{  const. } )
\end{align}
with the effective mass $m^* \equiv \hbar^2 / 2 t a^2$.  
Hence, in the continuum limit, the vector-potential dependent phase in our tight-binding Hamiltonian is equivalent to the usual minimal coupling Hamiltonian $\hat{H}_q$ in Eq.~(\ref{eq:HMinimal}) describing a charged particle in a magnetic field.  

\subsection{Basic methods for realizing synthetic gauge fields via driving}
\label{subsec:RealizingGaugeFields}

With this understanding in hand, we can return to our problem:  we would like non-trivial phases in our hopping Hamiltonian of 
Eq.~(\ref{eq:HTB}) without having to use charged particles and external magnetic fields.  The basic approach will be to obtain these phases by using nonlinearity and external driving (or time modulation) of our system.  To make the basic ideas here clear, let's first start with a reduced setup having only two modes.  We thus wish to generate an effective Hamiltonian of the form
\begin{equation}
	\hat{H}_{\rm eff} = 
		- t \left( e^{i \tilde{\phi} } \hat{a}^\dagger_2 \hat{a}_1 + h.c. \right)
		\label{eq:HPhaseTwoSites}
\end{equation}
with a non-zero, controllable hopping phase $\tilde{\phi}$.  There are two basic approaches to achieving this using driving.

\subsubsection{Coupling modulation}

Suppose we start with two modes that are non-resonant (i.e.~$\omega_1 \neq \omega_2$), and modulate the beam splitter coupling between them harmonically in time.  This is described by the Hamiltonian:
\begin{equation}
	\hat{H}_{cm}(t) = \omega_1 \hat{a}^\dagger_1 \hat{a}_1 + \omega_2 \hat{a}^\dagger_2 \hat{a}_2
		+ 2 \tilde{t} \cos( \omega_D t + \phi ) \left(\hat{a}^\dagger_2 \hat{a}_1 + h.c. \right)
		\label{eq:HCouplingMod}
\end{equation}
Let's move to a rotating frame generated by the unitary
\begin{equation}
	\hat{U}(t)= \exp \left[ i \left( \omega_1 \hat{a}^\dagger_1 \hat{a}_1 + \omega_2 \hat{a}^\dagger_2 \hat{a}_2 \right) t \right]
\end{equation}
In the new frame, the transformed wavefunctions are $| \psi'(t) \rangle = \hat{U}(t) | \psi(t) \rangle$, and they obey the time-dependent Schr\"{o}dinger equation $i \hbar \frac{d}{dt} | \psi'(t) \rangle = \hat{H}^\prime_{cm}(t) | \psi'(t) \rangle$
generated by the transformed Hamiltonian $\hat{H}'(t)$.  This is given by
\begin{align}
	\hat{H}^\prime_{cm}(t) & \equiv 
			\hat{U}(t) \hat{H}(t) \hat{U}^\dagger(t) + i \left( \frac{d}{dt} \hat{U} \right) \hat{U}^\dagger  \\
		& = 
			2 \tilde{t} \cos( \omega_D t + \phi ) 
				\left(\hat{a}^\dagger_2 \hat{a}_1 e^{i (\omega_2 - \omega_1)t}  + h.c. \right)
\end{align}		
We next pick the modulation frequency $\omega_D$ to be equal to the difference of the resonance frequencies of the two modes, i.e.~$\omega_D = \omega_2 - \omega_1$:
\begin{align}
	\hat{H}^\prime_{cm}(t) 
		& = 
			\tilde{t}  \left( e^{-i \phi} \hat{a}^\dagger_2 \hat{a}_1 + h.c. \right)
			+ \tilde{t}   
				\left(  e^{i \phi} \hat{a}^\dagger_2 \hat{a}_1 e^{2 i (\omega_2 - \omega_1)t}  + h.c. \right) 
\end{align}		
Finally, we further specialize to the situation where the coupling amplitude $\tilde{t}$ is much, much smaller than the frequency difference of the two modes:  $\tilde{t} \ll | \omega_2 - \omega_1 |$.  In this limit, the last bracketed term in the above Hamiltonian is highly non-resonant can be safely dropped (as in perturbation theory, it would yield small contributions controlled by the small parameter  $\tilde{t} /  | \omega_2 - \omega_1 |$).  This is nothing but the standard rotating-wave approximation (RWA).  

Making the RWA, we finally obtain an effective time-independent Hamiltonian that has the desired form of Eq.~(\ref{eq:HPhaseTwoSites}):  a beam splitter coupling with a controllable hopping phase.  The phase here is directly determined by the phase of the coherent sinusodial coupling modulation in the original time-dependent Hamiltonian.
At a heuristic level, it is useful to consider this final Hamiltonian as describing a three-wave mixing process where 
 a ``photon" from the classical modulation tone at frequency $\omega_D$ is either absorbed or emitted to facilitate resonant tunneling between modes $1$ and $2$.  

In a system of just two resonators,  the tunneling phase in our Hamiltonian could always be gauged away. However, the same modulation strategy can be directly generalized to lattices of 3 or more resonators to generate non-trivial phases and effective Aharanov-Bohm fluxes:  one just modulates each link of interest at the difference of the relevant resonance frequencies.  For example, consider a lattice with three sites described by 
Eq.~(\ref{eq:HTB}), with the replacements:
\begin{equation}
	t_{21} \rightarrow 2 \tilde{t} \cos \left(\omega_{21} t + \phi_A \right), \, \, \, \,
	t_{32} \rightarrow 2 \tilde{t} \cos \left(\omega_{32} t + \phi_B \right),  \, \, \, \,
	t_{13} \rightarrow 2 \tilde{t} \cos \left(\omega_{13} t + \phi_C \right)
	\label{eq:RingPhases}
\end{equation}
where $\omega_{ij} = \omega_i - \omega_j$.  Following the same steps as above, in the rotating frame (and after a rotating wave approximation), we obtain a time-independent tight binding Hamiltonian where the phases of each link are set by $\phi_A, \phi_B$ and $\phi_C$ respectively.  We can now define a gauge invariant ``synthetic flux" $\Phi = \phi_A + \phi_B + \phi_C$.  Such a phase cannot be eliminated by a local gauge transformation. 

One might still worry that our synthetic gauge flux ultimately relies on controlling the relative phases between modulation tones at different frequency, something that might seem to be ill defined.  We can express this worry more formally: each of the modulation phases $\phi_j$ 
in Eq.~(\ref{eq:RingPhases})
depends on our choice for the zero of time $t=0$.  If we shift the origin of time $t \rightarrow t + \tau$, then clearly these phases also change:
$\phi_A \rightarrow \phi_A + \omega_{21} \tau$, $\phi_B \rightarrow \phi_B + \omega_{32} \tau$, 
$\phi_C \rightarrow \phi_C + \omega_{32} \tau$.  Each of these phases is thus indeed sensitive to how exactly one decides to define the instant $t=0$.  This sensitivity is however not true for the loop flux $\Phi$:  it is independent of $\tau$ for the simple reason that $\omega_{21} + \omega_{32} + \omega_{13} = 0$.  We are left with the conclusion that the gauge-invariant synthetic gauge flux in our final time-independent Hamiltonian coincides with the single ``total" modulation phase in our time-dependent Hamiltonian that is defined independently of a specific choice of the zero of time.  

Before moving, we make an important note:  in practice, the coupling modulation strategy described here corresponds to using a parametric nonlinearity involving an auxiliary mode, which is driven strongly and hence treated classically.  To be concrete, let's return to our two mode problem. In general, the time-dependent two-mode Hamiltonian in 
Eq.~(\ref{eq:HCouplingMod}) is an approximation to a nonlinear system having one or more auxiliary modes that are driven.  For simplicity, consider the case where there is only a single auxiliary mode, and where the interaction and driving terms have the form
\begin{equation}
	\hat{H}_{\rm int} = g \left( \hat{b} + \hat{b}^\dagger \right) 
		\left( \hat{a}^\dagger_2 \hat{a}_1 + h.c. \right)
		+ \left( i f_D e^{-i \omega_D t} \hat{b}^\dagger + h.c. \right).
\end{equation}
The Hamiltonian has the form of a three-wave mixing (or $\chi_2$) style nonlinearity (amplitude $g$), where the auxiliary mode $b$ controls the tunneling of photons from mode $1$ and $2$.  Further, $f_D$ describes a simple linear drive on this auxiliary mode.  To obtain our coupling modulation Hamiltonian, we work in the usual limit where $g$ is weak and the drive $f_D$ is strong.  The equation of motion determining the average amplitude of mode $b$ is:
\begin{equation}
	\frac{d}{dt} \langle \hat{b} \rangle = 
		\left( -i \omega_b - \kappa_b/2 \right) \langle \hat{b} \rangle
		+ f_D e^{-i \omega_D t} + g (.....)
\end{equation}
Here $\kappa_b$ is the damping rate of the auxiliary mode.  
In the weak $g$ limit of interest, we can solve this equation for the steady state behaviour of  $\langle \hat{b}(t) \rangle$ ignoring the $g$ term:  
$\langle \hat{b}(t) \rangle = \bar{b} e^{-i \omega_D t}$, with $\bar{b} = f_D / ( -i (\omega_D - \omega_b) + \kappa_b / 2)$.  If we now replace $\hat{b} \rightarrow \bar{b} e^{-i \omega_D t}$ in our interaction Hamiltonian, we recover the coupling modulation Hamiltonian of Eq.~(\ref{eq:HCouplingMod}) with $\tilde{t} = g |\bar{b}|$, and the phase $\phi$ determined by the phase of $\bar{b}$.  Three-wave mixing Hamiltonians like this are common in many settings, e.g.~optomechanical setups \cite{Fang2017}, where a mechanical mode can modulate tunneling between two photonic modes.    A similar strategy for obtaining effective modulated-coupling beam-splitter Hamiltonians can be achieved starting with four-wave mixing Hamiltonians, as is commonly done in Josephson junction circuits (see e.g.~ \cite{Kamal2011,Abdo2013,Sliwa2015}).

\subsubsection{Frequency modulation}

We next consider an alternate means for obtaining synthetic gauge fields, where instead of modulating the couplings between modes, we instead modulate the resonance frequencies of each mode.  For simplicity, we again start with a simple two mode system.  Our starting time-dependent Hamiltonian now has the form:
\begin{equation}
	\hat{H}_{fm}(t) = 
		\left[ \omega_1 + A_1 \cos ( \Omega_1 + \phi) \right] \hat{a}^\dagger_1 \hat{a}_1 + 
		\omega_2 \hat{a}^\dagger_2 \hat{a}_2
		- t \left( \hat{a}^\dagger_2 \hat{a}_1 + h.c. \right)
\end{equation}
Similar to the treatment of coupling modulation, let's make a unitary transformation to eliminate the on-site time-dependent terms.  This is achieved via the unitary:
\begin{equation}
	\hat{U}(t) = 
		\exp \left[ i \omega_2 t \hat{a}^\dagger_2 \hat{a}_2 \right]
		\exp \left[ i \left( \omega_1 t + \frac{A_1}{\Omega_1} \sin( \Omega_1 t + \phi)  \right)
				\hat{a}^\dagger_1 \hat{a}_1 \right]
\end{equation}
The Hamiltonian in the rotating frame then corresponds to a time-dependent effective coupling:
\begin{equation}
	\hat{H}^\prime_{fm}(t) = 
		-t \left[
			\left( \exp[ i (\omega_2 - \omega_1) t ] 	
			\exp[ -i (A_1 / \Omega_1) \sin (\Omega_1 t + \phi)  ] \right) \hat{a}^\dagger_2 \hat{a}_1 + h.c. \right] 
\end{equation}

Next, recall the Jacobi-Anger identity:
\begin{equation}
	e^{-i z \sin \theta} = 
		\sum_{n = -\infty}^{\infty} J_n[z] e^{- i n \theta}
\end{equation}
where $J_n[z]$ is the $n$th Bessel function.  We see that the frequency modulation of mode $1$ results in a complicated modulation of the tunneling between mode 1 and mode 2, involving all harmonics of $\Omega_1$.  We now pick this modulation frequency so that only the first harmonic ($n=1$) results in a resonant tunneling process:  $\Omega_1 = \omega_2 - \omega_1$.  We further assume that all remaining terms can be safely neglected within the rotating wave approximation (i.e. they have a large frequency detuning / oscillation frequency compared to their amplitude).  Within this approximation, we again obtain a time-independent effective Hamiltonian where the phase of our modulation controls the effective phase of the hopping matrix element:
\begin{equation}
	\hat{H}^\prime_{fm} \simeq
		- t J_1 \left[ \frac{A_1}{\omega_2 - \omega_1} \right] 
		\left(
			e^{-i \phi} \hat{a}^\dagger_2 \hat{a}_1 + h.c. 
		\right)
\end{equation}

As usual, with just two modes, the phase $\phi$ can be gauged away and is not of any particular interest.  The minimal case for something interesting involves a ring of three modes, with the tunnel phases encoding a non-trivial flux.  The above frequency modulation strategy could be applied in this case.  Each mode has a time-dependent resonance frequency:
\begin{equation}
	\omega_j(t) = \omega_j + A_j \cos( \Omega_j t + \phi_j ) 
	\,\,\,\,\,\,\, (j=1,2,3)
\end{equation}
We consider a situation where the static, unmodulated frequencies of the three modes are all distinct, and e.g.:
\begin{align}
	\omega_2 & = \omega_1 + \Omega_1 \\
	\omega_2 & = \omega_3 + \Omega_2 \\
	\omega_3 & = \omega_1 + \Omega_3 
\end{align}
With these choices, we see that each mode-to-mode hopping process is made resonant using just one of the three frequency modulations $\Omega_j$.  An analogous derivation to the two mode case (and use of the rotating wave approximation) then yields an effective time-independent Hamiltonian that has the form of Eq.~(\ref{eq:HTB}) with eaching hopping phase controlled by the phase of one of the three modulation tones:
\begin{equation}
	t_{21} \propto e^{-i \phi_1}, \,\,\,\,\,
	t_{23} \propto e^{-i \phi_2}, \,\,\,\,\,
	t_{31} \propto e^{-i \phi_3}.
\end{equation}
This tight binding Hamiltonian corresponds to a gauge-invariant loop flux $\Phi = \phi_1 - \phi_2 - \phi_3$.  As before, one can easily check that this corresponds to a combination of modulation phases that is invariant under time-translation $t \rightarrow t + \tau$.

The idea of modulating on-site energies or resonant frequencies has been used in many systems to generate synthetic gauge fields; perhaps the best known examples come from cold atom systems (see e.g.~\cite{Aidelsburger2018}).  In cases where there is a choice, the coupling modulation approach of the previous subsection is usually preferable, as the requirements for the rotating wave approximation to be valid are less severe.  For the frequency-modulation scheme discussed here, there are in general many, many unwanted resonant sidebands that must be neglected.  This often limits one to extremely low modulation amplitudes.


\section{Dissipation and effective non-Hermitian dynamics for non-reciprocity}

Having introduced the notion of a synthetic gauge flux, we next would like to understand how these can directly lead to truly non-reciprocal scattering.  We will see that simply having a non-trivial set of phases is not enough:  one also needs dissipation to enter the system in just the correct manner.  As we will see, this can be compactly described by an effective non-Hermitian Hamiltonian that describes the propagation of particles within our system subject both to dissipation and the synthetic gauge field.  

\subsection{Basic model: three-site dissipative ring}
\label{subsec:BasicModel}

We will establish these ideas by analyzing the simplest setting where they arise:  a ring comprised of three photonic cavities, where photons can hop from mode to mode in the presence of a synthetic flux.  We work in a gauge where this phase is uniform on all three bonds, yielding a Hamiltonian:
\begin{equation}
	\hat{H}_3 = 
		-t \left[
			e^{-i \phi} \left(
				\hat{a}^\dagger_2 \hat{a}_1 + \hat{a}^\dagger_3 \hat{a}_2 + \hat{a}^\dagger_1 \hat{a}_3
			\right) + h.c.
		\right]
	\label{eq:HRing}
\end{equation}
The effective Aharanov-Bohm flux here (associated with traversing the ring once) is $\Phi = 3 \phi$; this phase cannot be gauged away unless $\Phi = n \pi$ for some integer $n$.  In what follows, we will always take $t > 0$ without loss of generality.  

\begin{figure}[t]
    \centering
    \includegraphics[width=0.7\textwidth]{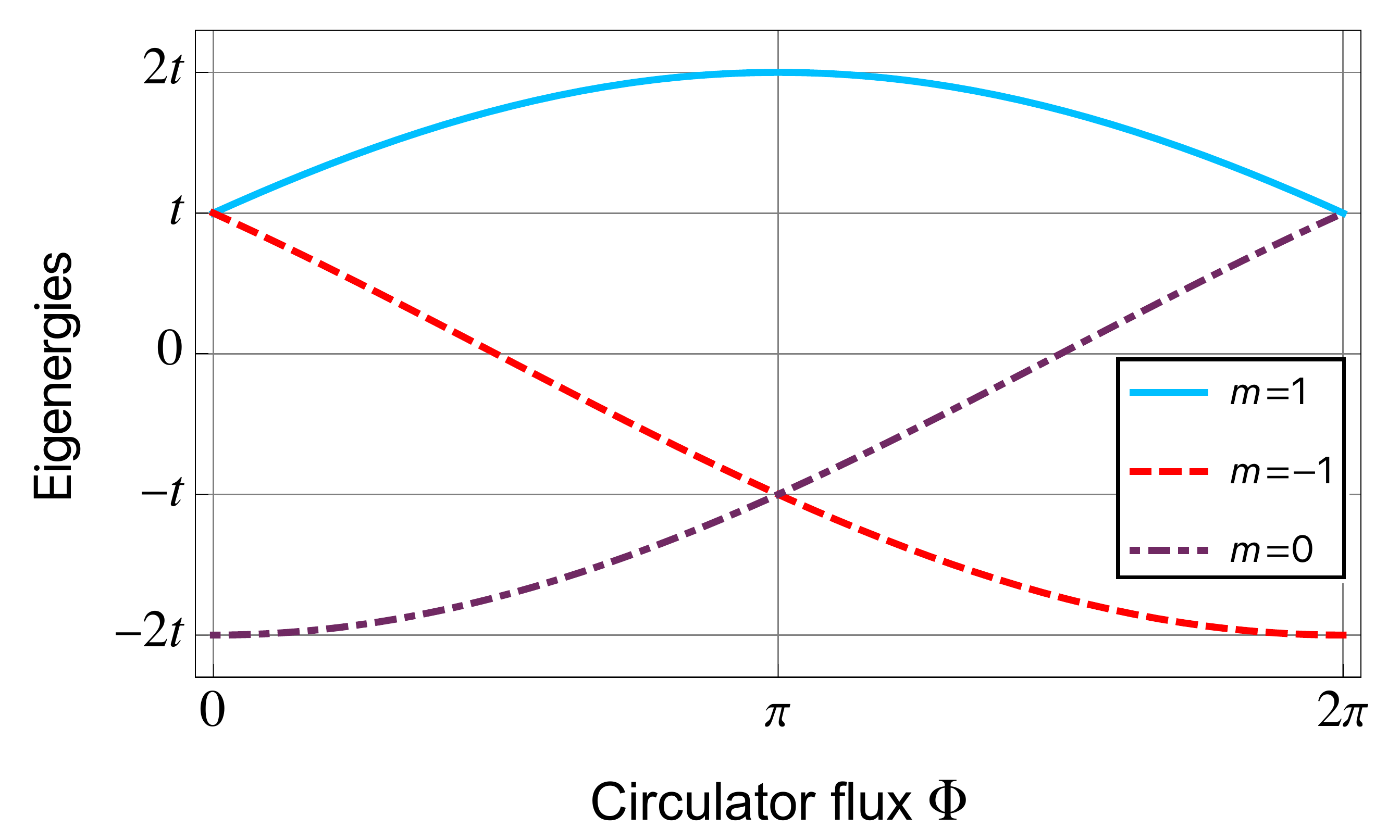}
    \caption{
        Energy eigenvalues of the three-cavity ring (c.f.~Eq.~(\ref{eq:HRing})), as a function of the synthetic gauge flux $\Phi$ threading the loop.  The energies are labelled according to the wavector $k_m = m 2 \pi / 3$ of the state; $m$ also corresponds to the quantized angular momentum of the eigenstate.  At the trivial values of the flux (i.e.~$\Phi = 0,\pi,2\pi$), the spectrum has a two-fold degeneracy reflecting the presence of time-reversal symmetry.  
    }
    \label{fig:EnergiesVsFlux}
\end{figure}

The above Hamiltonian has translational invariance and is thus easy to diagonalize in terms of plane waves.  The single-particle eigenstates are labelled by the wavevectors $k_m = \frac{2 \pi}{3} m$ ($m=-1,0,1$) with corresponding wavefunctions:
\begin{equation}
	| k_m \rangle = 
		\frac{1}{\sqrt{3}} \sum_{j=1}^3 e^{i k_m j} |j \rangle
\end{equation}
where $|j \rangle \equiv \hat{a}^\dagger_j | {\rm vac} \rangle$ corresponds to a ``position eigenket", i.e. a single photon localized on mode $j$.  One finds that the corresponding energy eigenvalues are
\begin{equation}
	\Omega_m = -2 t \cos( k_m + \phi) = -2 t \cos (k_m + \Phi/3 ).
\end{equation}
The second quantized Hamiltonian can thus be written as
\begin{equation}
	\hat{H}_3 = \sum_{m=-1,0,1} \Omega_m \hat{b}^\dagger_m \hat{b}_m
\end{equation}
with
\begin{equation}
	\hat{b}^\dagger_{m} = \frac{1}{\sqrt{3}} \sum_{j=1}^3 e^{i k_m j} \hat{a}^\dagger_j
\end{equation}
Note that we can interpret the $m$ label as indexing the quantized angular momentum associated with photons propagating along the ring either clockwise or anti-clockwise:  $m=0$ corresponds to zero angular momentum, while the eigenstates $m = \pm 1$ correspond to modes with one unit of angular momentum. 

Shown in Fig.~\ref{fig:EnergiesVsFlux} are the energy eigenstates of the ring as a function of the flux $\Phi$.  A few key things to take note of:
\begin{itemize}
	\item For ``trivial" values of the flux that can be gauged away (i.e.~$\Phi = 0, \pi, 2 \pi, ....$), we always have a degeneracy in the spectrum between two energy eigenvalues.  For $\Phi = 0$ this is between the $m = \pm 1$ modes, whereas for $\Phi = \pi$, it is between $m=0$ and $m=1$ modes.  
	\item The spectrum is identical for $\Phi = 0$ and $\Phi = 2 \pi$.  Increasing $\Phi$ by $2 \pi$ essentially adds one unit of angular momentum to each eigenstate.  Hence, as we vary $\Phi$ from $0$ to $2 \pi$, the $m=-1$ state is mapped to the $m=0$ state, the $m=0$ state is mapped to the $m=+1$ state, and the $m=1$ state is mapped to the $m=-1$ state.    
	\item Values of $\Phi$ away from these special trivial points break the degeneracy of the spectrum.  The spectral degneracy is maximally broken when $\Phi$ is a half-integer times $\pi$, as at these points, the levels are uniformly spaced.  For example, at $\Phi = \pi/2$, the three energy levels are $\Omega_m = t (- \sqrt{3},0,\sqrt{3} )$ for $m=0,-1,1$ respectively.  
\end{itemize}

\begin{figure}[t]
    \centering
    \includegraphics[width=0.7\textwidth]{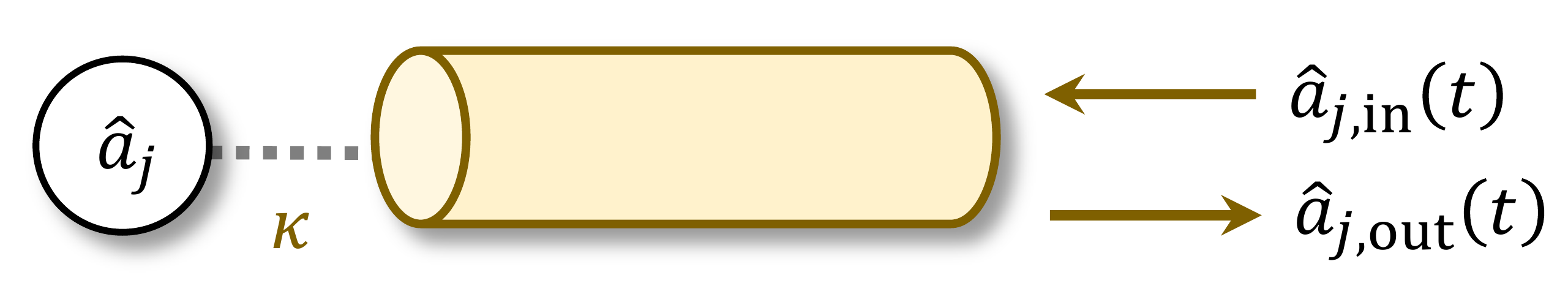}
    \caption{
       Schematic showing a single cavity (lowering operator $\hat{a}_j$) coupled to an input-output waveguide or transmission line.  $\hat{a}_{j,{\rm in}}$ encodes the the amplitude of signals and noise incident on the cavity, whereas $\hat{a}_{j,{\rm out}}$ encodes the amplitude of signals and noise leaving the cavity.
    }
    \label{fig:InputOutput}
\end{figure}

Our goal is to use this simple three site ring to build a scattering device having non-reciprocal properties.  To that end, we will couple each of our cavity modes $j=1,2,3$ to an input/output waveguide (see Fig.~\ref{fig:InputOutput}).  We will treat this waveguide coupling using the standard equations of input-output theory; for readers not familiar with this, see e.g.~\cite{Gardiner00,ClerkRMP} for pedagogical introductions.  Note that if we don't worry about noise terms, this treatment is identical to standard coupled mode theory equations as described in many engineering textbooks.  Allowing each cavity mode to be coupled to a waveguide at a rate $\kappa$, the Heisenberg-Langevin equations equations take the form:
\begin{equation}
	\frac{d}{dt} \hat{a}_j(t) = 
		- i [ \hat{a}_j(t), \hat{H}_3 ] - \frac{\kappa}{2} \hat{a}_j(t) - \sqrt{\kappa} \hat{a}_{j,{\rm in}}(t)
\end{equation}
Recall that the operators $\hat{a}_{j,{\rm in}}(t)$ describe both signals and noise incident on cavity $j$ through its coupling waveguide; both act as a source term in the above equation and directly act like a drive on cavity $j$.  Note that this operator has units of $1 / \sqrt{{\rm time}}$, as its corresponding number operator describes a photon number flux.    In a similar fashion, the operator
$\hat{a}_{j,{\rm out}}(t)$ describes signal and noise propagating outwards from cavity $j$ in its coupling waveguide.  Input output theory tells us that this output field is given by:
\begin{equation}
	\hat{a}_{j,{\rm out}}(t) = \hat{a}_{j,{\rm in}}(t) + \sqrt{\kappa} \hat{a}_j(t)
\end{equation}
The first term here describes ``promptly reflected" photons (i.e.~photons reflected at the entrance of the cavity), whereas the second term describes the emission from cavity $j$ into the waveguide.  

Next, suppose we apply a coherent drive on each of the three cavities through their respective waveguides, with different amplitudes, but all with the same frequency $\omega$.  This corresponds to
\begin{equation}
	\langle \hat{a}_{j,{\rm in}}(t) \rangle = \bar{a}_{{\rm in}, j} e^{-i \omega t}
\end{equation}
As our system of equations is completely linear, one can directly solve for the correspond amplitudes of the output field in each input-output waveguide.  One finds generally that
\begin{equation}
	\langle \hat{a}_{j,{\rm out}}(t) \rangle = \bar{a}_{{\rm out}, j} e^{-i \omega t}
\end{equation}
with a linear relation between these output amplitudes and the input amplitudes:
\begin{align}
	\bar{a}_{{\rm out}, j} & = 
		\sum_{j=1}^3 s_{jj'} [\omega] \bar{a}_{{\rm in}, j'} 		
\end{align}
Here $s_{jj'}[\omega]$ is the $3 \times 3$ scattering matrix that describes the scattering of both signals and noise off our systems of cavities.  While one could explicitly solve the equations of motion to obtain the elements of $s$, to make the physics clearer, we will use the general form of the solution that would be valid for any quadratic, photon-number conserving Hamiltonian for the three cavities:
\begin{align}
		s_{jj'}[\omega] & = 
			\delta_{jj'} - i \kappa G^R[j,j'; \omega]
			\label{eq:smatrixGR}
\end{align}
We have introduced here the retarded Green's function of our lattice, $G^R[j,j'; \omega]$.  This Green's function is a susceptibility, which tells us the photon amplitude induced on site $j$ if we drive the system at frequency $\omega$ on site $j$.  More usefully for us, it describes the propagation of photons {\it within the lattice.}.  Specifically, it gives the probability amplitude associated with the propagation of photons injected on site $j'$ to site $j$.  In general, the Kubo formula tells us:
\begin{equation}
	G^R(i,j;t) \equiv -i \theta(t) \langle [ \hat{a}_i(t), \hat{a}^\dagger_j(0) ] \rangle 
\end{equation}

As we have a single particle problem, the Green's function can also be constructed from the resolvant operator associated with the $3 \times 3$ matrix $H$ describing the first quantized version of our Hamiltonian, i.e.
\begin{equation}
	 G^R[j,j'; \omega] = 
	 	\left[
	 	\frac{1}{(\omega + i \kappa/2) \bm{I}_3 - \bm{H}} \right]_{jj'}
		\equiv
	 	\left[
	 	\frac{1}{(\omega ) \bm{I}_3 - \bm{H}_{\rm eff}} \right]_{jj'}		
		\label{eq:GRResolvant}
\end{equation}
where $\bm{I}_3$ is the identity matrix.  In the second line, we have introduced the effective $3 \times 3$ non-Hermitian Hamiltonian matrix $\bm{H}_{\rm eff} $ which encodes both the Hermitian coupling between the modes, as well as the the tendency of photons to leak out of the lattice into the waveguides (via the effective imaginary on-site energies $\propto \kappa$):
\begin{equation}
	\bm{H}_{\rm eff} = \bm{H} - i (\kappa/2) \bm{I}_3
		=
		\left(\begin{array}{ccc}
			-i \kappa/2 & -t  e^{i \phi} & -t  e^{-i \phi} \\
			-t e^{-i \phi} & -i \kappa/2 & -t  e^{i \phi} \\
			-t  e^{i \phi} & -t  e^{-i \phi} & -i \kappa/2
		\end{array}\right)
		\label{eq:NonHermMatrix}
\end{equation}

  To obtain an intuitive understanding of the scattering, we will express $G^R$ in terms of the energy eigenstates of our system:
\begin{equation}
	G^R[j,j'; \omega] = 
		\frac{1}{3} \sum_{m=-1}^1
		\frac{ e^{i k_m (j-j')} }{ \omega - \Omega_m + i \kappa/2} 
		\label{eq:GRpoles}
\end{equation}
We have a simple pole associated with each system eigenmode.  Heuristically, a photon injected on site $j'$ can propagate via any of the three eigenmodes in the system.  Each term describes the amplitude associated with one of these possibilities.  The final amplitude involves the coherent sum of the three possibilities, with the attendant possibilities of constructive and destructive interference.    

\subsection{Tuning flux and dissipation to achieve directional propagation within the ring}

With this general picture of scattering in hand, we can finally step back and ask: what exactly do we want of $s$?   For concreteness, let's try to engineer the simplest kind of non-reciprocal scattering matrix that encodes a definite directionality.  We will pick two ports in our system (say $j=1,2$) and try to construct the scattering matrix of an isolator at some frequency $\omega$:  signals at this frequency can be transmitted from $1 \rightarrow 2$ but not from $2 \rightarrow 1$.  We thus want to understand how we can tune parameters to achieve:
\begin{equation}
	s_{21}[\omega] = 0,  \,\,\,\,  s_{12}[\omega] \neq 0
\end{equation}
From our general expression above, this immediately yields a constraint on the Green's function:
\begin{equation}
	G^R[2,1; \omega] = 0,  \,\,\,\,   G^R[1,2; \omega] \neq 0
	\label{eq:GRNonRecipCondition}
\end{equation}
While going between these two conditions seems trivial, conceptually it represents a significant difference:  in the last condition, 
\emph{we are now solely focused on the propagation of photons within the lattice}.  Specifically, we want zero amplitude for propagation from $1$ to $2$, but a non-zero amplitude for the reverse process.  We stress that the propagation within the lattice (and any emergent non-reciprocity) is something that can be understood solely in terms of the non-Hermitian Hamiltonian $H_{\rm eff}$ introduced in Eq.~(\ref{eq:GRResolvant}).  

It is not surprising to guess that achieving the above directionality condition will require tuning the effective flux $\Phi$ in our Hamiltonian appropriately; this flux controls Green's function in Eq.~(\ref{eq:GRpoles}) solely through the dependence of the mode energies.  One can immediately check that if $\Phi = 0$, then the non-reciprocity condition of Eq.~(\ref{eq:GRNonRecipCondition}) cannot be fulfilled, as in this case, we necessarily have $G^R[2,1; \omega]  = G^R[1,2; \omega]$; this follows directly from the fact that $\Omega_1 = \Omega_{-1}$ when $\Phi = 0$.  Directionality will thus require a non-zero, non-trivial value of $\Phi$.  While one could reduce this to a purely algebraic exercise, we want a simple intuitive way of understand if and how one can find a magic value for $\Phi$.  As we now explain, there is indeed a simple principle at play here:  \emph{destructive interference}.  


\begin{figure}[t]
    \centering
    \includegraphics[width=0.7\textwidth]{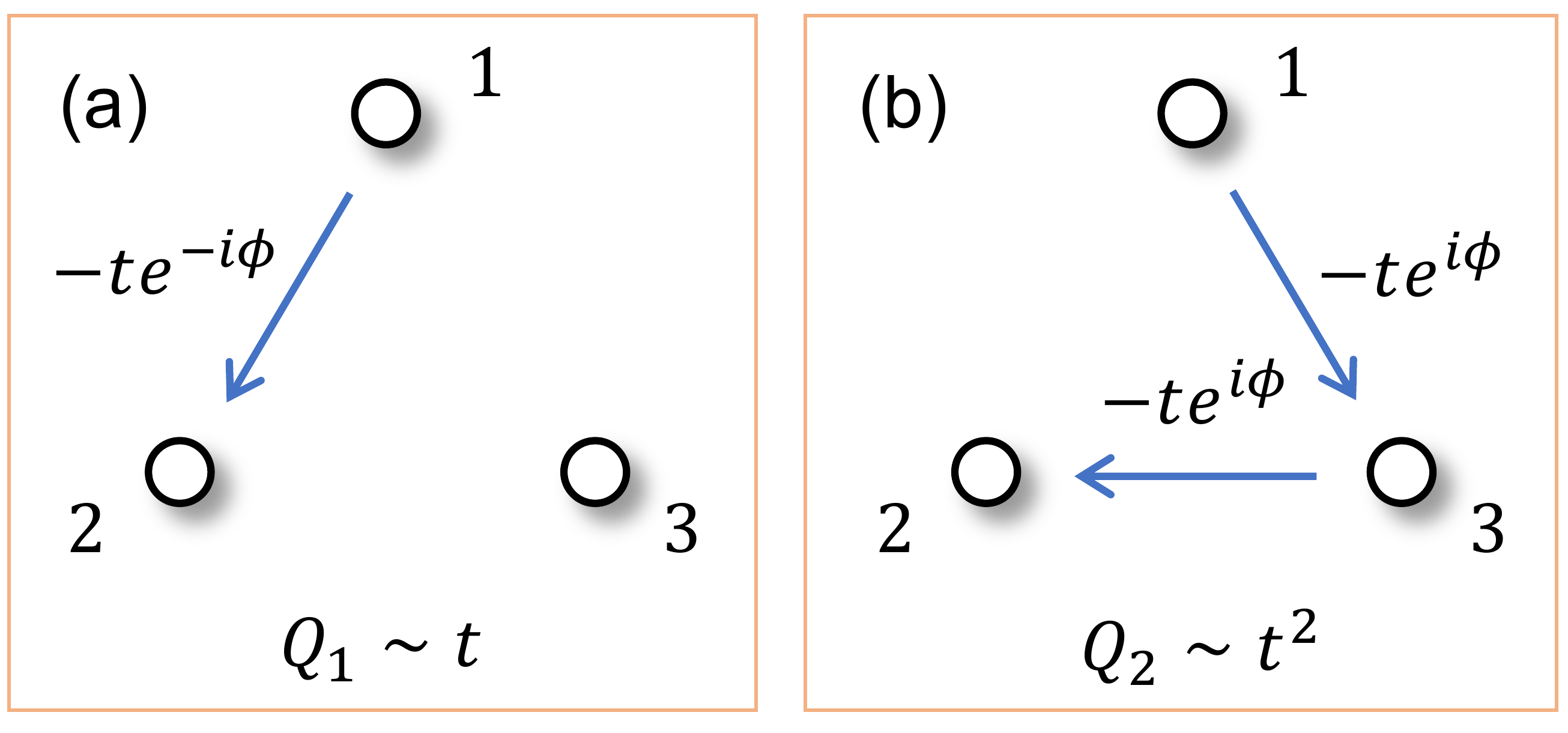}
    \caption{
        Figure showing the two simplest trajectories allowing propagation from from site 1 to 2.  Panel (a) shows a counter-clockwise trajectory whose probability amplitude is $\propto t$, whereas panel (b) shows a clockwise trajectory whose amplitude is $\propto t^2$.  The interference between these trajectories can be controlled by varying both the phase $\phi$ associated with the synthetic flux piercing the ring, the hopping amplitude $t$ and the damping rate $\kappa$ of mode 3 (which shows up in the energy denominator associated with the amplitude of process $Q_2$). 
    }
    \label{fig:SimpleTrajectories}
\end{figure}

Let's try to understand the value of the propagation amplitude encoded in $G^R(j,j';\omega)$ in terms of trajectories, i.e. different paths that could take the photon from site $j'$ to site $j$.  Formally, we can obtain such a picture by expanding $G^R$ in powers of the hopping $t$.  Shown in Fig.~\ref{fig:SimpleTrajectories} are the simplest trajectories that would take a photon initially on site $1$ to site $2$.  The counter-clockwise (CCW) trajectory involves a single hop and is labelled $Q_1$, while the the clockwise trajectory (CW) involves two hops and is labelled $Q_2$.  We can easily calculate the contribution of each of these processes to $G^R$.  We first define the unperturbed Green function $G_0$; this is the amplitude associated with residing on any given site in the absence of hopping, and is given by
\begin{equation}
	G_0 = \frac{1}{\omega + i \kappa/2}
\end{equation}
The amplitude for trajectory $Q_1$ is then:
\begin{equation}
	Q_1 = 
		G_0 \cdot \left(-t e^{-i \phi} \right) \cdot G_0
\end{equation}
Reading this equation from right to left, the first factor of $G_0$ is associated with starting on site 1, the bracketed factor is the clockwise hopping, and the last $G_0$ factor is associated with the last site.  This term corresponds to the first term in a Dyson series expansion of the full Green's function, where we view all of the hopping terms as a perturbation.  

In a similar fashion, the amplitude for the clockwise (CW) trajectory $Q_2$ is:
\begin{equation}
	Q_2 = 
		G_0 \cdot \left(-t e^{i \phi} \right) \cdot G_0 \left(-t e^{i \phi} \right) \cdot G_0
\end{equation}
This trajectory involves two hopping events, hence the two factors of $t$.  It also involves CW hopping, hence the phase factor for each hopping is $ e^{i \phi}$, and not 
 $e^{-i \phi}$ like we had in the CCW trajectory in $Q_1$.  Finally, the extra factor of $G_0$ compared to the $Q_1$ expression can be associated with the energy denominator we'd expect for a process in second order perturbation theory.  
 
 The processes $Q_1$ and $Q_2$ are the only contributions to $G^R[2,1; \omega]$ to order $t^2$.  Let's now enforce our directionality condition $G^R[2,1; \omega]=0$ to order $t^2$, which amounts to $Q_1 + Q_2 = 0$.  Substituting in the above expressions, this condition becomes:
\begin{equation}
	t e^{-i \phi}   =  \left(-t e^{i \phi} \right) \cdot G_0 \left(-t e^{i \phi} \right)  
\end{equation}
Some algebra lets us re-write this as a condition on the gauge invariant loop flux $\Phi$:
\begin{equation}
	e^{i \Phi} = \frac{\omega + i \kappa/2}{t}
\end{equation}
Hence, our first requirement for non-reciprocity (no 1 to 2 propagation) reduces to the two conditions:
\begin{align}
	\tan \Phi & =  \kappa / 2 \omega 
		\label{eq:NRCondPhi} \\
	\kappa/2 & = \sqrt{t^2 - \omega^2}
		\label{eq:NRCondKappa}  
\end{align}
Note crucially that these conditions require \emph{both} setting the value of the synthetic flux $\Phi$ as well as tuning the value of $\kappa$, i.e. tuning the strength of the dissipation induced in the cavities by the coupling to the waveguides.  In the simple case of $\omega = 0$ (i.e.~resonant driving of the cavities), the conditions reduce to $\kappa = 2 t$, and $\Phi = \pi/2$ (i.e.~a value of the flux that maximally breaks time reversal symmetry).   

\begin{figure}[t]
    \centering
    \includegraphics[width=0.7\textwidth]{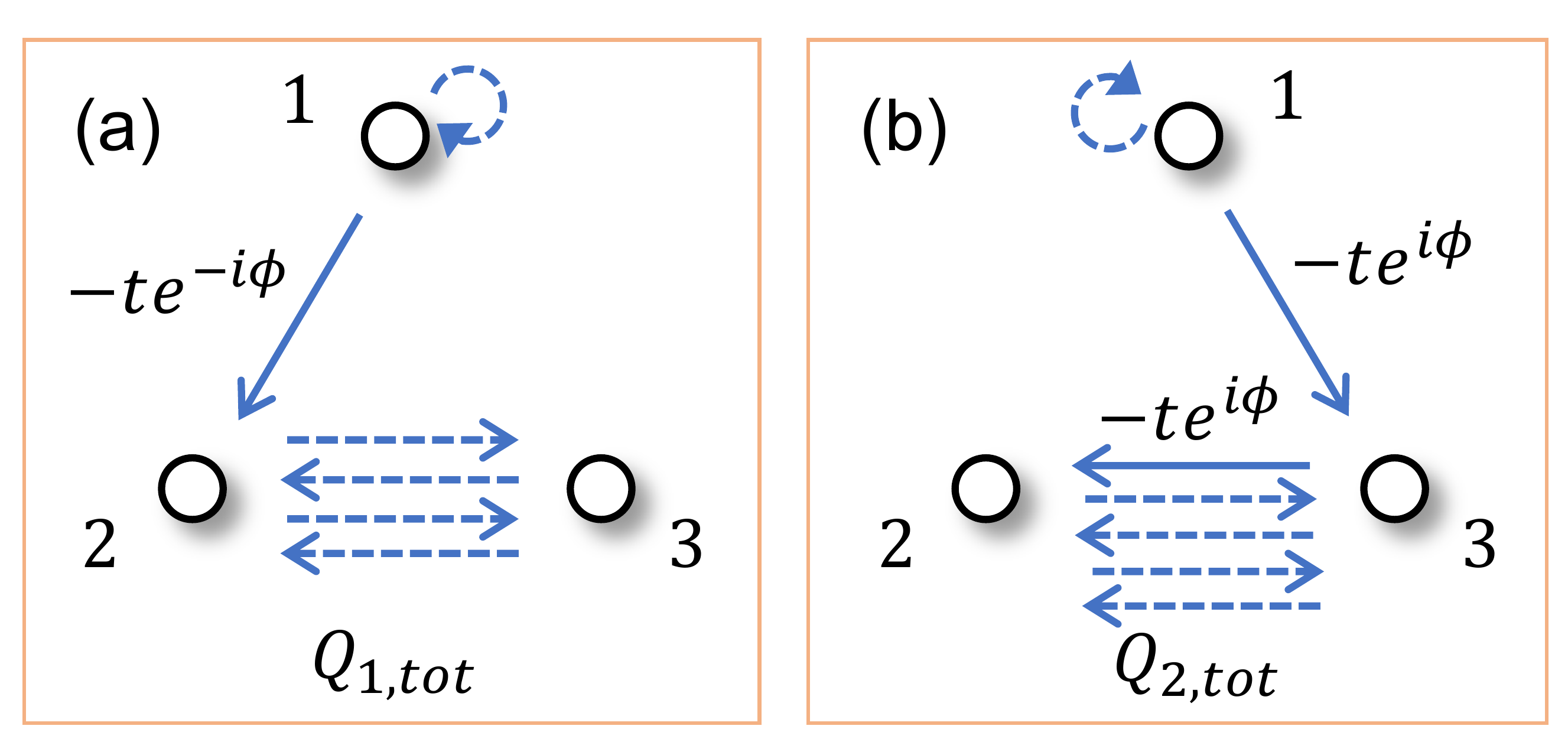}
    \caption{
       Schematic showing more complicated trajectories that contribute to the overall propagation from cavity 1 to cavity 2.  We can organize all of these into two categories, with respective amplitudes $Q_{1,tot}$ and $Q_{2,tot}$ (see text).
    }
    \label{fig:HigherOrderTrajectories}
\end{figure}

One might worry that the above conditions for cancelling $1 \rightarrow 2$ propagation are only valid for very small $t$, as they are based on a perturbative argument.  Surprisingly, this is not the case:  the same conditions ensure $G^R[2,1;\omega] = 0$ to all orders in $t$.  One can again see this from an intuitive argument
(see Appendix \ref{app:HigherOrderInterference} for a more rigorous formulation of the same argument).  Consider all trajectories that start on site $1$ and end on site $2$.
 We can partition these trajectories into two sets (see Fig.~\ref{fig:HigherOrderTrajectories}):
\begin{itemize}
	\item In the first set of trajectories a) the particle starts at $1$ then returns to $1$ via some arbitrary trajectory, then b) then particle hops to $2$, 
	then c) the particle hops in an arbitrary way back and forth between $2$ to $3$ before finally returning to $2$  
	Let's call the amplitude for all of these processes (which has contributions at all orders in $t$) $Q_{1,tot}$. 
	 Note that the probability amplitude for starting on site $j$, hoping in an arbitrary way, then returning to site $j$ is given by $G^R[j,j;\omega]$.  It then follows that:
	\begin{equation}
		Q_{1,tot} = Z[2,2;\omega] (-t e^{-i \phi} ) G^R[1,1; \omega]  \label{eq:Q1tot}
	\end{equation} 
	Here $Z[2,2;\omega]$ denotes the amplitude for all trajectories that start and end on $2$ with multiple hops to $3$.  
	Note that this trajectories are higher-order versions of the $Q_1$ process we discussed before.  
	\item The second set of trajectories are a similar generalization of the $Q_2$ process.  These are trajectories where
	a) the particle starts at $1$ then returns to $1$, then b) then particle hops to $3$ and then to $2$ 
	then c)  the particle hops in an arbitrary way back and form from $2$ to $3$ before finally returning to $2$ .  The probability amplitude $Q_{2,tot}$ for these trajectories is given by
	\begin{equation}
		Q_{2,tot} = Z[2,2;\omega] (-t e^{i\phi} ) G_0 (-t e^{i \phi} ) G^R[1,1; \omega] \label{eq:Q2tot}
	\end{equation} 
	Again, note the similarity to $Q_2$.
\end{itemize}
Considering now all trajectories, we have (to all orders in $t$) $G^R(1,2;\omega) = Q_{1,tot} + Q_{2,tot}$.  It is easy to see that the this sum is proportional to $Q_1 + Q_2$, and hence the \emph{identical} conditions in Eq.~(\ref{eq:NRCondPhi}) and (\ref{eq:NRCondKappa}) ensure that this will vanish.  Our destructive interference condition thus holds to all orders in the hopping $t$.  


The above conditions of course only ensure half of what we want (namely zero amplitude for propagation from site $1$ to site $2$).  We also want to ensure that amplitude for the reverse process, propagation from $2$ to $1$ has a non-zero amplitude.  One can again analyze this amplitude to order $t^2$ by expanding $G^R(1,2;\omega)$.   We again obtain a process that is first order in $t$ (amplitude $\tilde{Q}_1$), and a process that is second order in $t$ (amplitude $\tilde{Q}_2$).  It is easy to see (or explicitly verify) that these amplitudes are given respectively by substituting $\Phi \rightarrow -\Phi$ in the expressions for $Q_1$ and $Q_2$.  By then looking again at Eq.~(\ref{eq:NRCondPhi}):  if we pick a non-zero value of flux $\Phi$ that causes $G^R(2,1;\omega)$ to vanish via destructive interference, then the amplitude for the reverse process $G^ R(1,2;\omega)$ cannot be zero.  The same conclusion holds if we include terms to all orders in $t$ (as we did above).    Eqs. (\ref{eq:NRCondPhi}) and (\ref{eq:NRCondKappa}) thus yield the behaviour we are after:  propagation is allowed from $2$ to $1$, but not from $1$ to $2$.    Even at this level, we can draw some important conclusions:
\begin{itemize}
	\item Achieving non-reciprocal propagation in the lattice involves both tuning the synthetic flux to a non-trivial value, \emph{as well as} having the correct value of dissipation, i.e. value of $\kappa$.  
	\item If we are interested in non-reciprocity at zero frequency, then the synthetic flux must be $\Phi = \pi/2$, corresponding to a maximum breaking of time reversal symmetry.  
\end{itemize}

Finally, one might think that directional propagation within our lattice is possible even if $\kappa = 0$, if we work at some $\omega \neq 0$.  In this case, Eq.~(\ref{eq:NRCondPhi}) tells us that $\Phi = 0$ is needed to have $G^R(1,2;\omega)$ be zero.  But by the above argument, this will also cause the  
$G^R(1,2;\omega)$ to be zero!  There is thus no directionality here: we have cancelled hopping both directions (at this frequency) between sites $1$ and $2$.  This emphasizes a crucial point: non-reciprocal propagation within a lattice requires both a synthetic gauge field and non-zero dissipation.  
Before ending this subsection, we wish to emphasize that the basic picture of directionality arising from a specific kind of tailored destructive interference is also a crucial ingredient in the graph-theory approach to non-reciprocal linear scattering devices introduced in Ref.~\cite{Ranzani2015}. 

\subsection{From directional internal propagation to directional external scattering}

We now return to our original goal: ensuring directional scattering of waves incident on our three site ring.  We focus on $\omega = 0$.  From Eqs.~(\ref{eq:NRCondPhi}) and (\ref{eq:NRCondKappa}), we see that achieving directional propagation from $2 \rightarrow 1$ (and not in the reverse direction) requires tuning $\kappa = 2 t$ and $\Phi = \pi/2$.  This immediately implies:
\begin{equation}
	G^R(2,1;0) = 0 \,\,\, \implies s_{21}[0] = 0
\end{equation}
where we have used the general expression for the scattering matrix given in Eq.(\ref{eq:smatrixGR}).  

This is of course only part of what we would like for an ideal isolator.  We would additionaly want that $s_{12}[0] = 1$ (i.e.~unitary transmission in the forward direction of operation) and $s_{11} = 0$ (no reflections of signals incident on the input port $1$).  Let's first focus on this second condition, which requires
\begin{equation}
	s_{11}[0] \equiv 1 - i \kappa G^R[1,1;0] = 0
	\label{eq:s11}
\end{equation}
While we could obtain (as always) $G^R$ by simply doing the matrix inversion in Eq.~(\ref{eq:GRResolvant}), we will instead use a more intuitive argument.  Note that $G^R[1,1;0]$ describes the amplitude for all trajectories that start on site $1$ and then return to site $1$.  Apart from the trivial no-hopping process, there are three kinds of trajectories that contribute:
\begin{enumerate}
	\item Trajectories where the particle hops once from $1$ to $2$, then hops arbitrarily, then returns to $2$ then returns to $1$.  
	\item Trajectories where the particle hops once from $1$ to $3$, then hops arbitrarily, then returns to $3$ then returns to $1$.
	\item Trajectories where the particle hops once from $1$ to $3$, then hops arbitrarily returning to $3$, then hops to $2$, then returns to $1$
\end{enumerate}
Note that our specific tuning of $\kappa, \Phi$ makes the {\it total} amplitude for $1 \rightarrow 2$ propagation to vanish, and hence by symmetry, the total amplitude for 
$2 \rightarrow 3$ and $3 \rightarrow 1$ propagation also vanishes.  As such, the above three categories are the only kinds of trajectories we need to consider.  Given these three kinds of trajectories, and noting that $G^R(j,j;\omega)$ is the same for $j=1,2,3$ (by translational invariance), we can easily write down the expression for $G^R(1,1;\omega) \equiv \tilde{G}$:
\begin{equation}
	\tilde{G} = G_0 + 2 G_0 (-t) \tilde{G} (-t ) G_0  + e^{i \Phi} G_0 (-t) G_0 (-t) \tilde{G} (-t) G_0
\end{equation}
The second term on the RHS corresponds to processes 1 and 2, where as the last term corresponds to process 3.  

We can now solve this equation for $\tilde{G}$
\begin{align}
	\tilde{G}^{-1} & =   
		G_0^{-1} - 2 t^2 G_0 + e^{i \Phi} t^3 G_0^2 
			\nonumber \\ 
		& = 
			\frac{i \kappa}{2} - \frac{2 t^2}{i\kappa/2} + e^{i \Phi} \frac{t^3}{(i \kappa/2)^2} 
				\nonumber \\
		& = 
			\frac{\kappa}{2} \left( i - \frac{2}{i} - e^{i \Phi} \right) 
				\nonumber \\
		& = i \kappa
\end{align}
We have used the directionality tuning constraints above which set $\Phi = \pi/2$ and $t = \kappa/2$.  

Using this expression in Eq.~(\ref{eq:s11}), we find that, as desired, $s_{11}[0] = 0$.    Hence, by tuning the lattice dynamics to be non-reciprocal, we have automatically also forced the system to be impedance matched, i.e.~zero reflections at zero frequency on port 1.  By translational invariance, it also follows that $s_{22}[0] = s_{33}[0] = 0$.  Note that we can easily understand this lack of reflections in terms of a true impedance matching condition having been met.  Cavity one can be viewed as being coupled to two dissipative ports:  the input-output waveguide (coupling rate $\kappa$), and the effective dissipation coming from the coupling to cavity 3, which is damped.  Note that cavity 2 does not contribute because of the directionality.  The induced damping of cavity 1 via the coupling to cavity 3 is $\kappa_{eff} = 4 t^2 / \kappa = \kappa$.  Hence, the two ports coupled to cavity 1 have the same coupling rate, hence the lack of reflections.  

\begin{figure}[t]
    \centering
    \includegraphics[width=0.4\textwidth]{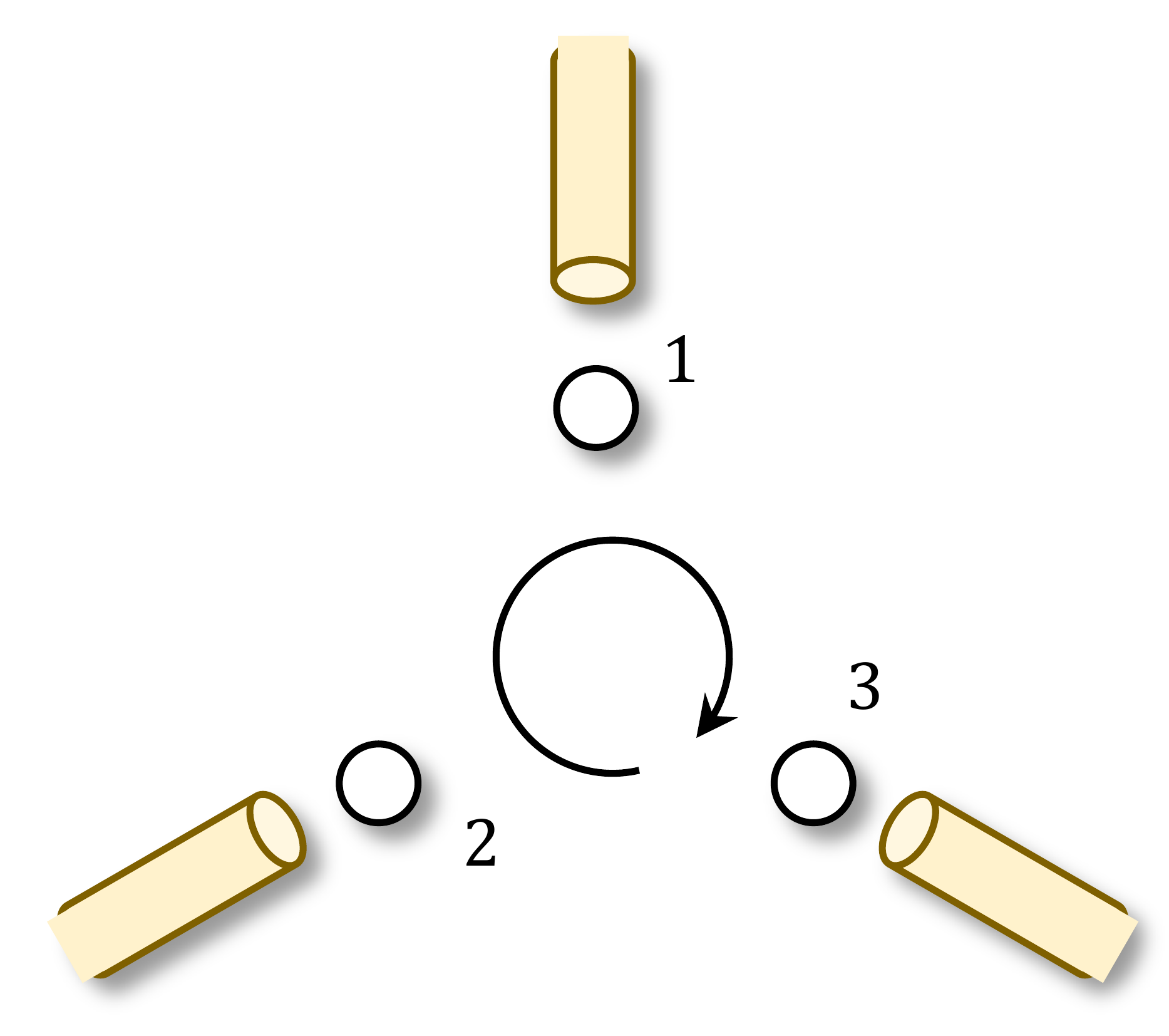}
    \caption{
        Schematic showing how our three-cavity ring (c.f.~Eq.~(\ref{eq:HRing})) when coupled to input-output waveguides (coupling rate $\kappa$) can function as a circulator of signals incident on the cavities at $\omega = 0$.  Achieving this circulator behaviour involves both tuning the synthetic gauge flux to $\Phi = \pi/2$, and also requires careful tuning of the amount of dissipation induced by the waveguides, i.e. $\kappa = 2 t$. 
    }
    \label{fig:Circulator}
\end{figure}

Turning to the full scattering matrix, note that translational invariance also tells us that all CCW scattering elements must be zero, and hence $s_{32}[0] = s_{13}[0] = 0$.  
We thus have that the zero frequency scattering matrix must have the form:
\begin{equation}
	\bm{s}[0] = 
		\left(\begin{array}{ccc}
			0 & z & 0 \\
			0 & 0 & z \\
			z & 0 & 0
		\end{array}\right)
\end{equation}
where $z$ is a complex number.  We must have $|z| = 1$ for the scattering matrix to be unity (one finds that $z=1$).  We thus have automatically also fulfilled our other requirement for a perfect isolator: perfect transmission in the forward direction.

While our goal was to construct an isolator, the translational invariance between all three modes in our system mean that we have in fact built a circulator, where scattering between ports is allowed in the forward, CW direction, but is zero in the CCW direction.  This scattering structure is depicted schematically in Fig.~(\ref{fig:Circulator}).

\subsection{Take home messages}

This simple three ring model has revealed several important lessons that are worth re-iterating:
\begin{itemize}
	\item  Achieving non-reciprocal scattering first requires achieving non-reciprocal propagation within the 3-site ring of cavities. 
	\item  This non-reciprocal propagation is a property of the non-Hermitian Hamiltonian $H_{\rm eff}$ which describes the propagation of photons between the cavities in the presence of the synthetic flux and dissipation (generated by the coupling waveguides (c.f.~Eq.~(\ref{eq:NonHermMatrix})).
	\item One cannot achieve non-reciprocal propagation without having dissipation
	\item Achieving non-reciprocity required matching the coherent and dissipative dynamics encoded in $H_{\rm eff}$, see Eq.~(\ref{eq:NRCondKappa}).   
\end{itemize}

\section{Effective models with dissipative interactions}

In the last section, we understood how one could use a simple ring of three cavities to construct an ideal circulator.  At the heart of this model was an effective $3 \times 3$ non-Hermitian Hamiltonian matrix that described propagation of photons in the ring subject to dissipation (loss) as well as a synthetic gauge field.  We saw that achieving directionality at a given frequency required tuning both the level of dissipation $\kappa$ as well as the value of the synthetic gauge field.  

In this section, we look at this three-ring system from a slightly different vantage point, one that will eventually let us understand how to generate non-reciprocal quantum interactions in a much more general way.  Recall that in the previous section, we were primarily interested in cavities $1$ and $2$, as we wanted a 2 port scattering device with the scattering matrix of an isolator.  As such, it is convenient and revealing to construct an effective model for just these two modes, where mode $3$ (and its coupling waveguide) are eliminated. As we now show, eliminating mode 3 results in the generating of an unusual dissipative interaction between modes $1$ and $2$ that plays a crucial role in establishing directionality.  

Let's start by generalizing the three-site ring model of the previous section.  We now take the hoppings to be asymmetric:
\begin{equation}
	|t_{12}| \equiv t,
		 \,\,\,\,\,\,\,\,  |t_{23}| = |t_{31}| = t'
\end{equation}
We also allow each mode $j$ to have a different damping rate $\kappa_j$ (i.e. coupling rate to its input-output waveguide).  Going forward, the goal is to get an isolator for ports 1 and 2; we don't care about anything having to do with mode 3 or its waveguide.  Further, we will assume that $\kappa_3 \gg t',t,\kappa_1,\kappa_2$.  This large damping rate will let us adiabatically eliminate mode 3 from our problem.

To proceed, we write the (Heisenberg-Langevin) equation of motion for mode 3:
\begin{equation}
	\frac{d}{dt} \hat{a}_3(t) = 
		-\frac{\kappa_3}{2} \hat{a}_3 + i t' \left( e^{-i \phi} \hat{a}_2(t) + e^{i \phi} \hat{a}_1(t) \right) - \sqrt{\kappa_3} \hat{a}_{3, {\rm in}}(t)
\end{equation}
We can Fourier transform this equation, and then solve for $\hat{a}_3[\omega]$:
\begin{equation}
	\hat{a}_3[\omega] = \frac{1}{-i \omega + \kappa_3/2} \left(
		i t' \left( e^{-i \phi} \hat{a}_2[\omega] + e^{i \phi} \hat{a}_1[\omega] \right) - \sqrt{\kappa_3} \hat{a}_{3, {\rm in}} [\omega]
		\right)
\end{equation}
We now assume that $\kappa_3$ is much larger than the frequencies $\omega$ that we care about (e.g. the bandwidth of signals that we will feed into our isolator).  We can thus ignore the $\omega$ in the denominator of the above expression.  Returning to the time domain, we see that this is an adiabatic approximation:  the heavily damped $\hat{a}_3(t)$ mode has an amplitude that is determined by the instantaneous value of the other two modes:
\begin{equation}
	\hat{a}_3(t) \simeq
		\frac{2 i t'}{\kappa_3} \left( e^{-i \phi} \hat{a}_2(t) + e^{i \phi} \hat{a}_1(t) \right) 
		- \frac{2}{\sqrt{\kappa_3}} \hat{a}_{3,in}(t)
\end{equation}		

Consider now the EOM for the $\hat{a}_1$ mode:
\begin{equation}
	\frac{d}{dt} \hat{a}_1(t) = 
		-\frac{\kappa_1}{2} \hat{a}_1 + i  \left( t e^{i \phi} \hat{a}_2(t) + t' e^{-i \phi} \hat{a}_3(t) \right)  + (...)
\end{equation}
where we do not write the noise terms explicitly.  Substituting in the value of $\hat{a}_3(t)$, we find:
\begin{equation}
	\frac{d}{dt} \hat{a}_1(t) \simeq 
		-\frac{1}{2} (\kappa_1 + \tilde{\kappa} ) \hat{a}_1 + i \tilde{t}_{12}  \hat{a}_2(t)    + (...)
\end{equation}
with
\begin{equation}
	\tilde{\kappa} = \frac{4 (t')^2}{\kappa_3} ,
	\,\,\,\,
	\tilde{t}_{12} = e^{i \phi} t + e^{-2 i \phi} \frac{2 i (t')^2}{\kappa_3} 
\end{equation}
Here $\tilde{\kappa}$ represents extra induced damping from mode $3$ on mode $1$ whereas $\tilde{t}_{12}$ 
 describes the modified effective hopping from mode $2$ to $1$ (where the second term is a virtual process going through mode 3).  

An analogous calculation for $\hat{a}_2$ yields:
\begin{equation}
	\frac{d}{dt} \hat{a}_2(t) \simeq 
		-\frac{1}{2} (\kappa_2 + \tilde{\kappa} ) \hat{a}_2 + i \tilde{t}_{21}  \hat{a}_1(t)    + (...)
\end{equation}
with $\tilde{\kappa}$ as above, and
\begin{equation}
	\tilde{t}_{21} = e^{-i \phi} t + e^{2 i \phi} \frac{2 i (t')^2}{\kappa_3} 
\end{equation}
Note crucially that the tunnel couplings between modes 1 and 2 in the above equations cannot correspond to a Hermitian beam-splitter Hamiltonian, as $\tilde{t}_{12} \neq \tilde{t}_{21}^*$.  We see that the contribution to these hopping matrix elements from mode 3 gives them a non-Hermitian structure.  Formally, we can write
\begin{equation}
	\tilde{t}_{12} = e^{i \phi} t + \tilde{t}_{\rm diss} \,\,\,\,  \tilde{t}_{21} = e^{-i \phi} t - \left( \tilde{t}_{\rm diss} \right)^*
\end{equation}
where $\tilde{t}_{\rm diss}$ describes a ``dissipative" tunneling coupling that is mediated by mode $3$.   Ignoring noise terms, we could describe this structure by writing our equations of motion in terms of a non-Hermitian effective Hamiltonian matrix $\bm{H}_{\rm eff,2}   $, i.e.
\begin{equation}
	\frac{d}{dt}
		\left(\begin{array}{c} 
		\hat{a}_1 \\
		\hat{a}_2 
	\end{array}\right) 
		=  -i \bm{H}_{\rm eff,2}   
			\left(\begin{array}{c} \hat{a}_1 \\
		\hat{a}_2 
	\end{array}\right) 
\end{equation}

with
\begin{equation}
	 \bm{H}_{\rm eff,2}  = 	
		\left(\begin{array}{cc}
			-i (\kappa_1 + \tilde{\kappa})/2 & e^{i \phi} t + \tilde{t}_{\rm diss} \\
			e^{-i \phi} t - \left( \tilde{t}_{\rm diss} \right)^* & -i (\kappa_2 + \tilde{\kappa})/2
		\end{array}\right)
		\label{eq:Heff2}
\end{equation}

Within this formulation, it becomes even easier to see how to have mode $1$ influenced by mode $2$, but not vice-versa:  we simply tune $\tilde{t}_{\rm diss}$ so that $\tilde{t}_{21} = 0$.  Because of the non-Hermitian structure, this does not simultaneously force $\tilde{t}_{12}$ to be 0 as well.   
Matching the magnitudes of the coherent and dissipative tunnelings, we find the condition:
\begin{equation}
	\left( \frac{\tilde{\kappa}}{2} \equiv \frac{2 (t')^2}{\kappa_3} \right) = t
\end{equation}
This is reminiscent of Eq.~(\ref{eq:NRCondKappa}) of the last section.  If this condition is met, we then have:
\begin{equation}
	\tilde{t}_{12/21} = e^{\pm i \phi} t \left( 1 + i e^{\mp 3i \phi} \right)
\end{equation}
Hence, if we pick $\Phi = 3 \phi = \pi/2$, we obtain $\tilde{t}_{21} = 0$ (i.e.~no tunneling from 1 to 2), whereas $\tilde{t}_{12} = 2 t e^{i \pi/4} $. This flux tuning
matches the condition in Eq.~(\ref{eq:NRCondPhi}).

While it seems like we have simply re-derived the results of the previous section, the approach here leads to an important physical interpretation that can be generalized to many different situations:
\begin{itemize}
\item Modes 1 and 2 are coupled together in two ways.  The first is the (Hermitian) Hamiltonian coupling described by the hopping matrix element $t$; we will often refer to this as a coherent coupling, as dissipation is not involved.
\item Modes 1 and 2 are also coupled to a common bath, i.e.~the highly damped mode $\hat{a}_3$.  This mode mediates a {\it dissipative} hopping interaction between modes 1 and 2
\item  The Hamiltonian interaction on its own is reciprocal; it allows hopping in both directions between modes 1 and 2.  The same is true for the purely dissipative hopping interaction.
\item  However, when we combine both of these kinds of hopping processes, we can get directionality.  We can have the two processes cancel for one direction of the hopping but not the other.
\end{itemize}
We thus have a very general recipe:  {\bf non-reciprocal interactions arise from the balancing of ``coherent'' and ``dissipative'' interactions}.  

The above discussion of mode-3 eliminated system has been phrased using equations of motion and an effective non-Hermitian Hamiltonian matrix. This is a convenient description for linear systems, and (without noise terms) has been used extensively in the study of classical non-Hermitian systems (where non-Hermiticity is usually obtained by applying incoherent loss and or gain).  For quantum systems that are not linear, the above formulation is not the most convenient.  Instead, it is useful to describe the dissipative interactions generated by the bath of interest (here mode 3) using a quantum master equation.

To that end, we can view the part of our original (Hermitian) Hamiltonian that involves mode 3 as a kind of system bath Hamiltonian:
\begin{equation}
	\hat{H}_{SB} = - t' \hat{a}^\dagger_3 \hat{z}_{\rm sys} + h.c., \,\,\,\,\,  \hat{z}_{\rm sys} = e^{-i \phi} \hat{a}_2 + e^{i \phi} \hat{a}_1
\end{equation}
Here, we interpret $\hat{a}_3^\dagger$ as an operator that creates an excitation in the bath, whereas $\hat{z}_{\rm sys}$ is the system operator that couples to bath.  To be explicit, we can create an excitation in the bath either removing a photon from mode 1 or 2 (with coherence and interference between these two possibilities).  

In the limit of large $\kappa_3$, we could now go through the standard steps of treating mode 3 as a reservoir where any created excitations decay quickly, and derive a quantum master equation, i.e.~an equation of motion for the reduced density matrix describing mode 3.    The steps in deriving the master equation are discussed in many standard references (see e.g.~\cite{Gardiner00}).  The resulting Lindblad master equation has the standard form:
\begin{equation}
	\frac{d}{dt} \hat{\rho} = 
		-i [ \hat{H}_{12} , \hat{\rho} ]
		+ \tilde{\kappa} \left( 
			\hat{z}_{\rm sys} \hat{\rho} \hat{z}_{\rm sys}^\dagger - \frac{1}{2} \{ \hat{z}_{\rm sys}^\dagger \hat{z}_{\rm sys}, \hat{\rho} \} \right)
	\label{eq:FullMaster}
\end{equation}
where
\begin{equation}
	\hat{H}_{12} = -t \left( e^{-i \phi} \hat{a}^\dagger_2 \hat{a}_1 + h.c. \right)
\end{equation}
The first term of the master equation describes the usual coherent (i.e. non-dissipative) evolution under the hopping Hamiltonian $\hat{H}_{12}$.  The remaining terms describe the dissipative effect of the reservoir.  Heuristically, there is a probability per unit time $\tilde{\kappa}$ of having a quantum jump where the system state suddenly changes from $| \psi \rangle$  to $\hat{z}_{\rm sys} | \psi \rangle$ due the creation of an excitation in the bath.  Note that we can make a gauge transformation $\hat{a}_1 \rightarrow e^{i \phi} \hat{a}_1$ to eliminate the phase from $\hat{H}_{12}$.  Further, the overall phase of $\hat{z}_{\rm sys}$ plays no role. We can thus re-write the master equation in a form where the non-trivial phase factor only appears in the dissipative part of the equation:
\begin{align}
	\frac{d}{dt} \hat{\rho} & = 
		-i [ -t \left(  \hat{a}^\dagger_2 \hat{a}_1 + h.c. \right), \hat{\rho} ]
		+ \tilde{\kappa} \left( 
			\hat{z} \hat{\rho} \hat{z}^\dagger - \frac{1}{2} \{ \hat{z}^\dagger \hat{z}, \hat{\rho} \} \right)
			\label{eq:MasterRingExplicit} \\
			\hat{z} & = 
				\hat{a}_2 + e^{i \Phi} \hat{a}_1
\end{align}
This master equation fully describes the possibly directional dynamics of hopping between our cavities 1 and 2 in the adiabatic limit of interest (where $\kappa_3$ is large).  The dynamics is fully directional when we fulfill the same tuning conditions listed above, i.e.
\begin{equation}
	 \tilde{\kappa} /2 = t, \,\,\,\,\,  \Phi = \pm \pi/2
\end{equation}
Here, the $+$ sign gives us $2 \rightarrow 1$ directionality, the $-$ sign $1 \rightarrow 2$ directionality.  One can easily confirm that equations of motions for the average values of $\hat{a}_1$ and $\hat{a}_2$ obtained from the master equation are completely consistent with the Heisenerg-Langevin equations we used above. 

Before leaving this simple model discussion, it is interesting to note that we can extract an effective non-Hermitian Hamiltonian from the master equation.  We can re-write it in the form
\begin{equation}
	\frac{d}{dt} \hat{\rho} = -i \left( \hat{H}_{\rm NH} \hat{\rho} - \hat{\rho}  \hat{H}^\dagger_{\rm NH} \right) + \tilde{\kappa} \hat{z} \hat{\rho} \hat{z}^\dagger 
	\label{eq:MasterHeff}
\end{equation}
with
\begin{align}
		\hat{H}_{\rm NH} & = -t \left(  \hat{a}^\dagger_2 \hat{a}_1 + h.c. \right) - i \frac{\tilde{\kappa}}{2} \hat{z}^\dagger \hat{z} \\
		& = 
			\left(-t - i \frac{\tilde{\kappa}}{2} e^{i \Phi} \right) \hat{a}^\dagger_2 \hat{a}_1	+
			\left(-t - i \frac{\tilde{\kappa}}{2} e^{-i \Phi} \right) \hat{a}^\dagger_1 \hat{a}_2	
			-  i \frac{\tilde{\kappa}}{2} \left( \hat{a}^\dagger_1 \hat{a}_1 + \hat{a}^\dagger_2 \hat{a}_2 \right)					
\end{align}
This corresponds to the non-Hermitian Hamiltonan we identified directly from the equations of motion, c.f. Eq.~(\ref{eq:Heff2}).  We thus see that our master equation describes both the non-reciprocal, non-Hermitian dynamics of photons in the ring, along with the correspond noise terms (i.e. the remaining terms in Eq.(\ref{eq:MasterHeff})).  Surprisingly, we will see that the above master equation can be generalized to describe non-reciprocal interactions that {\it cannot} be associated with the non-Hermitian Hamiltonian part of the master equation, $\hat{H}_{\rm NH}$.  Note that $\hat{H}_{\rm NH}$ plays a distinctive role when one ``unravels" the master equation in terms of quantum trajectories.  In this approach, one has a stochastic Schr\"{o}dinger equation for the system; averaging over this stochastic process results in the final master equation.  For these stochastic trajectories, $\hat{H}_{\rm NH}$ describes the evolution of the system in the absence of a quantum jump.  We thus see that in this case, the no-jump evolution in a stochastic trajectory is described by a non-Hermitian Hamiltonian whose form can be directional.  The ability to realize effective non-Hermitian Hamiltonians by monitoring a system and post-selecting to no-jump trajectories has been studied by several recent works (see e.g.~\cite{Murch2019}).

\section{General quantum model for non-reciprocal interactions}

In previous sections, we saw in detail how one-way,  non-reciprocal dynamics emerged in a simple model of a ring of three cavities, where each cavity was coupled to a waveguide.  We established that non-reciprocal scattering was directly connected to non-reciprocal propagation \emph{within} the ring, something that could be accomplished by designing a non-Hermitian Hamiltonian with both dissipation and a non-trivial synthetic gauge flux (c.f.~Eq.~(\ref{eq:NonHermMatrix})).  We also saw in the last section that we could obtain an effective description of the non-reciprocity between cavities 1 and 2 where the the third (highly-damped) cavity was adiabatically eliminated.  In the result effective model, directionality emerged as the balancing of a coherent Hamiltonian interaction (i.e. simple tunneling between modes 1 and 2), and a ``dissipative interaction" mediated by mode 3.  This latter interaction was described by a collective dissipation term in the quantum master equation (c.f.~Eq.~(\ref{eq:MasterRingExplicit})).  

In this section, we now show that the structure of the master equation in Eq.~(\ref{eq:MasterRingExplicit}) can be generalized to make {\it any} starting interaction between two quantum systems non-reciprocal.  This will allow us to establish a general recipe for designing non-reciprocal quantum interactions, in settings where it is impossible to reduce the dynamics to an effective non-Hermitian Hamiltonian.  Our discussion here follows Ref.~\cite{Metelmann2015}, but provides additional heuristic insights.

\subsection{Basic structure}

Consider a general situation where we have two quantum systems $1$ and $2$ (described by a tensor-product Hilbert space), that interact via a Hermitian Hamiltonian of the form
\begin{equation}
	\hat{H}_{\rm coh}(\lambda)= \frac{1}{2} \left( \lambda \hat{O}_1 \hat{O}_2 + \lambda^* \hat{O}_1^\dagger \hat{O}_2^\dagger   \right)
	\label{eq:HCoherent}
\end{equation}
Here, $\hat{O}_1$ is an operator acting on system 1, and $\hat{O}_2$ an operator acting on system 2.  This Hamiltonian describes a reciprocal interaction, in that the evolution of system $1$ will in general depend on what system $2$ is doing, and vice versa.  For convenience, we take these operators to be dimensionless, hence $\lambda$ has the units of frequency and controls the strength of the interaction.  

\begin{figure}[t]
    \centering
    \includegraphics[width=0.4\textwidth]{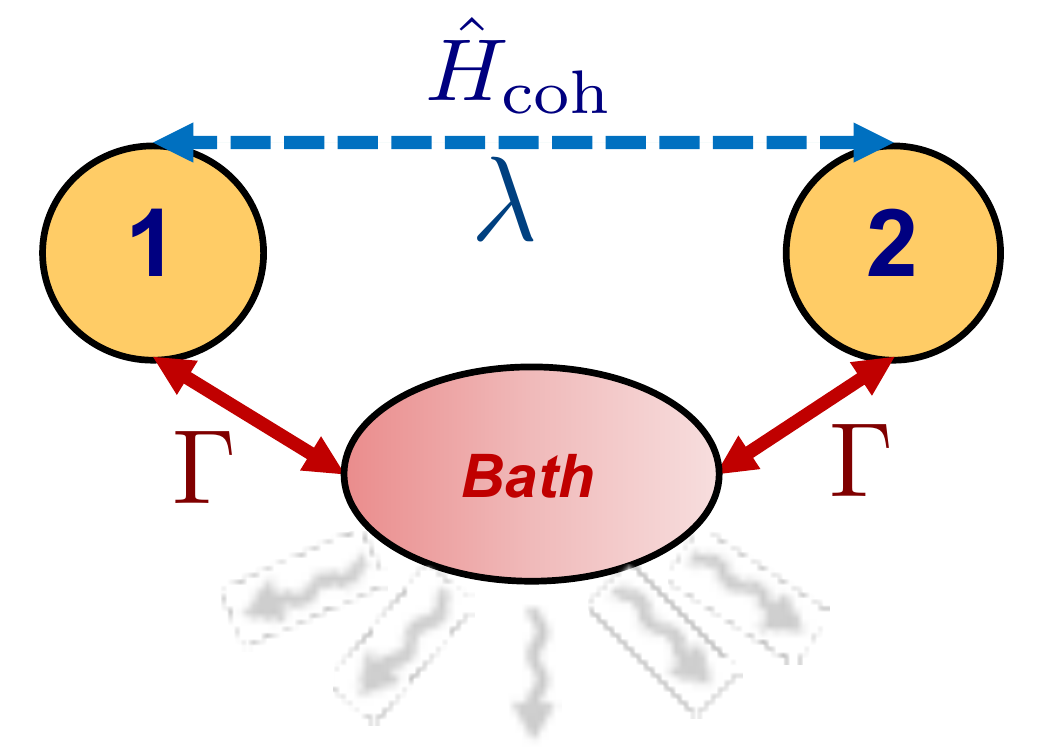}
    \caption{
        Schematic showing the basic setup for generating an arbitrary non-reciprocal interaction between two quantum systems $1$ and $2$.  We start with a coherent Hamiltonian interaction $\hat{H}_{\rm coh}$ between the two systems involving the product of operators $\hat{O}_1$ and $\hat{O}_2$ for each system, with an interaction strength $\lambda$.  By introducing correlated dissipation on both systems in just the right way (via coupling to a common reservoir at rate $\Gamma$), a non-reciprocal interaction can be achieved.   
    }
    \label{fig:DissNonRecipSchematic}
\end{figure}

The goal is to convert this interaction into a non-reciprocal interaction that is fully directional.  To accomplish this, we take the same approach as in the last section: we will achieve directionality by balancing the coherent interaction in $\hat{H}_{\rm coh}$ with a dissipative interaction mediated by a reservoir that couples to both system 1 and 2 in just the right way.  This dissipation interaction will be described by dissipative terms in a master equation analogous to the last terms in Eq.~(\ref{eq:FullMaster}).  We are thus led to consider a master equation of the form:
\begin{align}
	\frac{d}{dt} \hat{\rho} & = 
		-i [ \hat{H}_{\rm coh}(\lambda), \hat{\rho} ]
		+ \Gamma \mathcal{D}[ \hat{O}_1 - i e^{i \theta} \hat{O}_2^\dagger ] \hat{\rho}
\end{align}
where the dissipative superoperator $\mathcal{D}$ is defined as
\begin{equation}	
	\mathcal{D}[\hat{z}] \hat{\rho} \equiv \hat{z} \hat{\rho} \hat{z}^\dagger - \frac{1}{2} \{ \hat{z}^\dagger \hat{z}, \hat{\rho} \}
\end{equation}
Again, the last dissipative terms here correspond to coupling system 1 and 2 to a common reservoir that mediates a dissipative interaction.  Here $\Gamma$ controls the strength of this dissipative interaction, while the phase $\theta$ controls its form.  Also note the non-trivial phase factors in the dissipative terms:  as we will see, these encode an effective synthetic gauge flux, something that we saw was crucial to obtaining non-reciprocity.    

We now claim that by tuning $\Gamma$ and $\theta$ appropriately, we can achieve a fully directional interaction between systems $1$ and $2$.  To see whether this is possible, let's consider how the expectation value of system 1 and system 2 operators evolve under the above dynamics.  Take $\hat{A}_1$ ($\hat{A}_2$) to be an arbitrary system $1$ (system $2$) operator.  As they act on different systems, it is natural to take these operators to commute with one another
\footnote{
	The one exception to having system $1$ and $2$ operators commute would be the case where both systems are fermionic, meaning that $A_1$ and $A_2$ could anti-commute with one another.   One can confirm that our derivation of the equations of motion for averages remains unchanged in this case. }  .
Let's now calculate the equation of motion for their average values using the above master equation.  Consider first 
$\hat{A}_1$:
\begin{align}
	\frac{d}{dt}  \left \langle \hat{A}_1(t) \right \rangle & \equiv 
		\textrm{tr} \left( \hat{A}_1 \frac{d}{dt} \hat{\rho}(t) \right) \nonumber \\
%
	 & =
		\Gamma \textrm{tr}   \left( \hat{A}_1 \mathcal{D}[\hat{O}_1] \, \hat{\rho} \right)
		-i \, \textrm{tr} \left( 
			 \left[  \hat{A}_1, \frac{\tilde{\lambda}_{12}}{2} \hat{O}_1 \hat{O}_2 + \textrm{h.c.} \right] \, \hat{\rho} \right) 
			 \label{eq:A1AverageEOM}
\end{align}	
where we have defined
\begin{equation}
	\tilde{\lambda}_{12} = \lambda + \Gamma e^{-i \theta}
	\label{eq:LamTilde12}
\end{equation}
For $\Gamma = 0$ (i.e.~no dissipation), we have the expected coherent evolution generated by $\hat{H}_{\rm coh}$, i.e.~Eq.~(\ref{eq:A1AverageEOM}) reflects the usual Heisenberg equation of motion for $\hat{A}_1$.  Dissipation (i.e.~terms proportional to $\Gamma$) has two effects:  it adds purely local dissipative dynamics on system one, generating the first term on the RHS of Eq.~(\ref{eq:A1AverageEOM}), and also generates a dissipative interaction.  This dissipative interaction leads to an interaction that seems analogous to $\hat{H}_{\rm coh}$, and would seem to just modify the interaction strength from $\lambda \rightarrow \tilde{\lambda}_{12}$.  Concretely, as far as system $1$ is concerned, the evolution is indistinguishable from the master equation:
\begin{align}
	\frac{d}{dt} \hat{\rho} & = 
		-i [ \hat{H}_{\rm coh}(\tilde{\lambda}_{12}), \hat{\rho} ]
		+ \Gamma \mathcal{D}[ \hat{O}_1 ] \hat{\rho},
\end{align}
i.e.~the dissipative interaction looks like a modification of the Hamiltonian interaction strength.

Let's now repeat this exercise for an arbitrary system $2$ operator $\hat{A}_2$:
\begin{align}
	\frac{d}{dt}  \left \langle \hat{A}_2(t) \right \rangle & \equiv 
		\textrm{tr} \left( \hat{A}_2 \frac{d}{dt} \hat{\rho}(t) \right) 
		\nonumber \\
	 & =
		\Gamma \textrm{tr}   \left( \hat{A}_2 \mathcal{D}[\hat{O}_2^\dagger] \, \hat{\rho} \right)
		-i \, \textrm{tr} \left( 
			 \left[  \hat{A}_2, \frac{\tilde{\lambda}_{21}}{2} \hat{O}_1 \hat{O}_2 + \textrm{h.c.} \right] \, \hat{\rho} \right) 
			  \label{eq:A2AverageEOM}
\end{align}	
where we have defined
\begin{equation}
	\tilde{\lambda}_{21} = \lambda - \Gamma e^{-i \theta}
		\label{eq:LamTilde21}
\end{equation}
Again, as far as system 2 is concerned, it is as though we modified the Hamiltonian interaction strength from $\lambda \rightarrow \tilde{\lambda}_{21}$, and also introduced some puirely local dissipation one mode $2$.  

We now see from Eqs.~(\ref{eq:LamTilde12}) and (\ref{eq:LamTilde21}) that there is a crucial asymmetry:  if $\Gamma \neq 0$, then the effective interaction strength seen by system $1$, $\tilde{\lambda}_{12}$ can differ both in magnitude and phase from $\tilde{\lambda}_{21}$, the effective interaction strength seen by system $2$.  Our working definition of non-reciprocity here will be situations where these couplings differ in magnitude.  In particular, we can obtain a perfect non-reciprocal situation by tuning the dissipative couplings so that:
\begin{equation}
	\Gamma e^{-i \theta} = - \lambda
\end{equation}
In this case we have:
\begin{equation}
	\tilde{\lambda}_{12} = 0,  \,\,\,\, \tilde{\lambda}_{21} = 2 \lambda
\end{equation}
We thus achieve a perfectly non-reciprocal coupling between the two systems.  System $1$ is not influenced by system 2 at all, but system 2 is driven by system 1 in the same way as would be achieved with a Hamiltonian $\hat{H} = \hat{H}_{\rm coh}(2 \lambda)$.  If we instead replace $\lambda \rightarrow -\lambda$ in the above equation, we would obtain perfect non-reciprocity in the opposite direction.  Note that in the fully directional case, the isolated system does not experience any direct driving by the other system, but does feel some local dissipation whose strength is proportional to $\lambda$ (e.g.~the first terms on the RHS of Eqs.~(\ref{eq:A1AverageEOM}) and (\ref{eq:A2AverageEOM}).  The presence of this dissipation is an unavoidable consequence of trying to generate a strongly directional interaction, and mirrors what we found in our analysis of the three-site ring.  

 For completeness, we write the final directional master equation.  Redefining $O_1, O_2$ so that $\lambda$ is real and positive, it has the form
\begin{align}
	\frac{d}{dt} \hat{\rho} & = 
		-i \frac{\lambda}{2}  [ \hat{O}_1 \hat{O}_2 + \hat{O}_1^\dagger \hat{O}_2^\dagger  , \hat{\rho} ]
		+ \lambda \mathcal{D}[ \hat{O}_1 \mp  i \hat{O}_2^\dagger ] \hat{\rho}
	\label{eq:FinalGeneralDirectionalMEQ}
\end{align}
 where the upper (lower) sign corresponds to $1 \rightarrow 2$  ($2 \rightarrow 1$) directionality.  With this form, we explicitly can see the parallel to the simple three mode model studied in the previous, c.f.~Eq.~(\ref{eq:MasterRingExplicit}).  

We stress that the above master equation results in a fully nonreciprocal interaction between systems $1$ and $2$ no matter what the choice of the coupling operators $\hat{O}_1$ and $\hat{O}_2$.  Also note that there is no clever way to remove the explicit factor of $i$ in the dissipator above from the entire master equation:  if we try to redefine e.g. $\hat{O}_2$ to absorb this phase, it will then show up explicitly in the Hamiltonian.  We thus see that the basic ingredients needed for non-reciprocity in our simple three-site toy model (namely synthetic gauge fields and dissipation) also fuel this more general recipe.

\begin{figure}[t]
    \centering
    \includegraphics[width=0.4\textwidth]{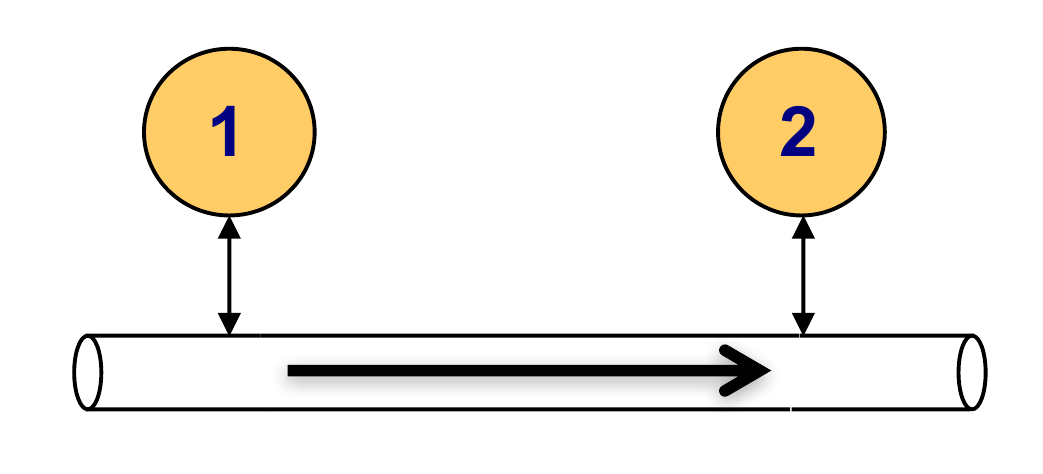}
    \caption{
        Schematic showing the basic setup of a cascaded quantum system, where two quantum systems interact via a unidirectional (i.e.~``chiral") waveguide.  In the Markovian limit, the interactions mediated between the two systems via the waveguide correspond exactly to our effective directional Lindblad master equation.    
    }
    \label{fig:CascadedSchematic}
\end{figure}

Eq.~(\ref{eq:FinalGeneralDirectionalMEQ}) has the form of a so-called \emph{cascaded quantum master equation}, which provides another way to understand its origin.  This is the effective master equation that emerges when two quantum systems are coupled to a unidirectional waveguide, such that e.g.~waves emitted from mode $1$ can only propagate towards mode $2$, and not-vice versa (see Fig.~\ref{fig:CascadedSchematic}).  In this case, both terms in the master equation (i.e.~the coherent and dissipative interactions) are generated by the directional waveguide.  Our discussion provides another way of thinking about the structure of this master equation without any recourse to a directional waveguide: the emphasis is on synthetic gauge fields and interference between coherent and dissipative interactions.  As we have already seen, this structure can be realized experimentally using external driving, as was discussed in Sec.~\ref{subsec:RealizingGaugeFields}.

\subsection{Important properties}

\subsubsection{Alternate realizations: asymmetric bath couplings}

Eq.~(\ref{eq:FinalGeneralDirectionalMEQ}) represents one concrete way to turn the basic Hamiltonian interaction in Eq.~(\ref{eq:HCoherent}) into something fully directional.  It however is not the only strategy.  The crucial part of our recipe was to balance a dissipative interaction against a coherent interaction.  The dissipative interaction formally corresponds to dissipative terms in our master equation that involve the product of $\hat{O}_1$ and $\hat{O}_2$.  As such, the prefactors of these operators need not be equal, as long as the product of prefactors remains the same.  More physically, this means that we can couple systems 1 and 2 asymmetrically to the bath.  This leads to a more general class of directional master equations, labelled by the asymmetry parameter $\eta > 0$:
\begin{align}
	\frac{d}{dt} \hat{\rho} & = 
		-i \frac{\lambda}{2}  [ \hat{O}_1 \hat{O}_2 + \hat{O}_1^\dagger \hat{O}_2^\dagger  , \hat{\rho} ]
		+ \lambda \mathcal{D}[ \eta^{1/2} \hat{O}_1 \mp  i  \eta^{-1/2}  \hat{O}_2^\dagger ] \hat{\rho}
	\label{eq:MEQAsymmetric}
\end{align}
One can easily check that for any value of $\eta >0$, this master equation is completely directional; this follows from the fact that the dissipative interaction is independent of $\eta$.  To see this, return to the general equations of motion of expectation values of system 1 and 2 operators.  One finds that the interaction terms in Eqs.~(\ref{eq:A1AverageEOM}) and (\ref{eq:A2AverageEOM}) are unchanged.  The only difference are the local dissipation terms in each equation:  in the $\hat{A}_1$, this local dissipation terms is now $\propto \eta$, whereas in the $\hat{A}_2$ equation is $\propto 1 / \eta$.  

This flexibility in achieving directionality is useful.  We know that non-reciprocity must come with an increase in local dissipation on each system.  By varying $\eta$, one can spread this cost unevenly between the two systems.  

\subsubsection{Conjugated scheme for non-Hermitian coupling operators}

In the general case where the operators $\hat{O}_j$ are both non-Hermitian, one might be puzzled at first glance by the asymmetry in the final master equation Eq.~(\ref{eq:FinalGeneralDirectionalMEQ}):  why is one of the coupling operators conjugated, and not the other?  This asymmetry also manifests itself in the local dissipation terms in the equations of motion for means, Eqs.~(\ref{eq:A1AverageEOM}) and (\ref{eq:A2AverageEOM}).   As it turns out, the other choice is also possible, i.e.
\begin{align}
	\frac{d}{dt} \hat{\rho} & = 
		-i \frac{\lambda}{2}  [ \hat{O}_1 \hat{O}_2 + \hat{O}_1^\dagger \hat{O}_2^\dagger  , \hat{\rho} ]
		+ \lambda \mathcal{D}[ \hat{O}_1^\dagger \mp  i \hat{O}_2 ] \hat{\rho}
\end{align}

This also generate fully directional behaviour with a direction depending on the sign.  The difference is now in the form of the local dissipation terms that act independently on each system.  

\subsubsection{Efffective non-Hermitian Hamiltonian}

Similar to what we did in Sec.~\ref{subsec:BasicModel}, we can extract the non-Hermitian Hamiltonian associated with our master equation in Eq.~(\ref{eq:FinalGeneralDirectionalMEQ}).  One finds:
\begin{align}
	\hat{H}_{\rm eff} & \equiv 
		\hat{H}_{\rm coh} - \frac{i}{2} \lambda
		\left(
			\hat{O}_1  \mp i \hat{O}_2^\dagger
		 \right)^\dagger 
			\left(
			\hat{O}_1 \mp i \hat{O}_2^\dagger \right)  \nonumber \\
		& =  \frac{\lambda}{2} \left( \hat{O}_1 \hat{O}_2 + \hat{O}_2^\dagger \hat{O}_1^\dagger \right)  
		- \frac{i}{2} \lambda \left( \hat{O}_1^\dagger \hat{O}_1 + \hat{O}_2\hat{O}_2^\dagger  \right)
			\pm \frac{1}{2} \lambda \left( \hat{O}_1 \hat{O}_2 - \hat{O}_2^\dagger \hat{O}_1^\dagger \right)
\end{align}

For $\hat{O}_1$ and $\hat{O}_2$ non-Hermitian, we clearly see that there is a strong asymmetry in this effective Hamiltonian between modes $1$ and $2$, as the dissipative terms cancel one of the two terms in the coherent Hamiltonian.  This matches what we found for the simple three-site ring in the previous section.  More strange is the case where both $\hat{O}_1$ and $\hat{O}_2$ are both Hermitian.  In this case, the last dissipative interaction term in $\hat{H}_{\rm eff}$ \emph{vanishes}.  This leads to an important case:  for the case where both system operators in the coupling are Hermitian, the non-reciprocity predicted by Eq.~(\ref{eq:FinalGeneralDirectionalMEQ}) cannot be understood in terms of an effective non-Hermitian Hamiltonian.   We will provide an alternate intuition for this case in the next section.


\section{Quantum non-reciprocity and quantum measuerment-plus-feedforward schemes}

We have now established a very general, quantum master equation recipe for generating arbitrary non-reciprocal interactions between two quantum systems.  We understood the basic mechanism as being the balancing of a coherent Hamiltonian interaction against a bath-mediated dissipative interaction.  In this section, we sketch a seemingly very different physical situation that leads to unidirectional dynamics:  measurement plus feedforward.  Surprisingly, we see that the standard theory of such a protocol (in the limit of a weak continuous measurement) leads again to an equation analogous fo Eq.~(\ref{eq:FinalGeneralDirectionalMEQ}).

\subsection{Quantum feedforward based on weak continuous measurements}

\begin{figure}[t]
    \centering
    \includegraphics[width=0.8\textwidth]{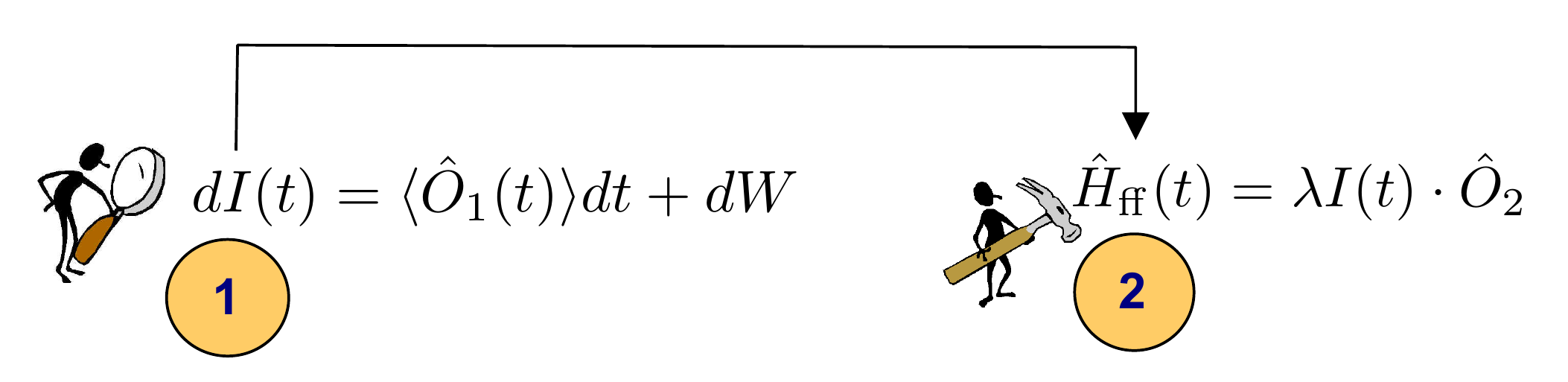}
    \caption{
       Schematic of a generic measurement plus feedforward scheme.  The observable $\hat{O}_1$ of system $1$ is continuously monitored, with the results of the measurement (encoded in the measurement record $I(t)$ used to drive system 2 via the operator $\hat{O}_2$.  The unconditional evolution under this scheme results in a purely directional interaction between the two systems.  In the limit where delay in applying the feedforward force can be ignored, the resulting evolution corresponds to our general non-reciprocal master equation (with the proviso that both coupling operators must be Hermitian). 
    }
    \label{fig:FeedforwardSchematic}
\end{figure}

We consider again two systems $1$ and $2$ that are not directly interacting (either via a direct Hamiltonian interaction, or via a common dissipative bath).  Instead, we analyze a situation where an observer makes a weak continuous measurement of an observable $\hat{A}_1$ in system $1$, and then uses the results of this measurement to drive system $2$ (via a forcing operator $\hat{F}_2$ (see Fig.\ref{fig:FeedforwardSchematic})  The theory of weak continuous measurements is treated in several places, see \cite{Steck2006} for an extremely clear pedagogical treatment.  The continuous classical measurement record is denoted $I(t)$ (i.e.~this could be an integrated homodyne signal or electrical current).  We assume that in each infinitesimal time interval $[t, t+ dt]$ the measurement record increases an amount $dI(t)$.  This increment has two terms: a piece that reflects the value of the measured observable $\hat{A}_1$, and a random noise amount $dW_t$:
\begin{equation}
	dI(t) = \sqrt{k} \langle \hat{A}_1(t) \rangle dt + dW_t
\end{equation}
$k$ here represents the strength of the measurement and has units of a rate, while $dW_t$ is a random variable, a Wiener increment.  It can be viewed as integrating white noise for a time $dt$, and satisfies $\overline{dW_t} = 0$, $dW_t^2 = dt$ (where the overline indicates a stochastic average, i.e. averaging over different measurement outcomes). The Wiener increments in different time intervals are completely uncorrelated, reflecting the fact that this is white noise.  In what follows, we drop the $t$ index on these Wiener increments.

We can now ask how the density matrix of the system $\hat{\rho}$ evolves in a particular run of the experiment, i.e.~conditioned on the measurement record $I(t)$.  The conditional master equation governing this situation is:
\begin{equation}
	d \hat{\rho} = \frac{k}{4} \mathcal{D}[ \hat{A}_1 ] \hat{\rho} dt +
		\frac{\sqrt{k}}{2} \left \{  \hat{A}_1 - \langle \hat{A}_1 \rangle , \hat{\rho} \right \} dW 
		 \equiv  \mathcal{L}_0 \hat{\rho}
\end{equation}
We stress that the $dW$ here is the same Wiener increment appearing in the expression for the measurement record.  The first term here represents the unconditional backaction of the measurement:  if we don't have access to the measurement record, then the only relevant backaction is the disturbance of quantities that do not commute with $\hat{A}_1$.  The second term in our equation represents a backaction on the system that is correlated with the measurement record.  This is often described as the  ``conditioning" of the system by the measurement, or an ``information-gain" backaction.  Heuristically, it is equivalent to updating a prior distribution in Bayesian statistics based on the acquisition of new information.  

We next imagine that the experimentalist uses the measurement record to apply a time-dependent generalized force on system $2$.  Letting $J(t) = (d/dt) I(t)$, we consider that this feedforward forcing is described by the Hamiltonian:
\begin{equation}
	\hat{H}_{\rm FB}(t) = 
		J(t - \tau) \sqrt{\gamma_{\rm ff}} \hat{F}_2
\end{equation}
Here $\hat{F}_2$ is some system-2 Hermitian operator, $\gamma_{\rm ff}$ is the feedforward strength (with units of rate) and $\tau > 0$ is the delay time associated with applying the feedforward.  

Let's now consider the evolution of the full conditional density matrix in the presence of feedforward, during an infinitesimal time interval $dt$.  Because of causality, we should first evolve the state to reflect the ``backaction" of the measurement of $\hat{A}_1$ during the interval, and then evolve it via the feedforward Hamiltonian $\hat{H}_{\rm FB}$ (which uses the measurement record acquired during the interval).  We thus have
\begin{align}
	\hat{\rho}(t+dt) & = 
		\mathcal{U}_{\rm ff}(t) \cdot \mathcal{U}_{\rm meas}(t) \cdot \hat{\rho}(t)
\end{align}  
where both $\mathcal{U}_{\rm meas}(t)$ and $\mathcal{U}_{\rm ff}(t) $ are superoperators.  
$\mathcal{U}_{\rm meas}(t)$ describes evolution due to measurement backaction, and for $dt \rightarrow 0$ can be written
\begin{equation}
		\mathcal{U}_{\rm meas}(t) \simeq  1 + \mathcal{L}_0(t) 
\end{equation}
In contrast, $\mathcal{U}_{\rm ff}(t) $ is the unitary evolution of the system under the feedforward Hamiltonian.  Defining the superoperator	
\begin{equation}
	\mathcal{M} \hat{\rho} \equiv
		-i [ \sqrt{\gamma_{\rm ff}} \hat{F}_2, \hat{\rho} ]
\end{equation}
we have
\begin{align}
	\mathcal{U}_{\rm ff}(t) & = \exp \left[ J(t-\tau) \mathcal{M} dt \right]  \\
		& \simeq
		1 + dI(t-\tau) \mathcal{M}  + \frac{1}{2} \mathcal{M}^2 dt
\end{align}  
In the last line, we have expanded the exponential to order $dt$, being mindful of the Ito rule $dW^2 = dt$.

Combining these terms and only keeping terms to order $dt$, we find in the limit $\tau \rightarrow 0^+$:
\begin{align}
	d \hat{\rho}(t) & = 
		\left(
			\mathcal{L}_0  + dI(t) \mathcal{M} + \frac{1}{2} \mathcal{M}^2 dt
		\right) \hat{\rho}(t) 
		+
		 \mathcal{M} \frac{\sqrt{k}}{2} 
			\left \{  \hat{A}_1 - \langle \hat{A}_1 \rangle ,  \hat{\rho}(t) \right \}  dt
\end{align}			
The last term here is the most interesting one.  It arises from {\it correlations} between the $dW$ noise in the feedforward evolution, and the $dW$ terms in the backaction (innovation) of the measurement propagator. 

Up until now, we have been looking at the conditional evolution of the system density matrix: how the system evolves given a particular measurement record $J(t)$.  We now consider the unconditional evolution, i.e. average over all measurement outcomes.  This is equivalent to averaging over the noise $dW$ i.e. setting $dW = 0$ in the above equation.  This yields:
\begin{align}
	d \hat{\rho}(t) & = \left[
		\frac{k}{4} \mathcal{D}[\hat{A}_1]   +  \sqrt{k} \langle \hat{A}_1 \rangle \mathcal{M} 
		+ \frac{1}{2} \mathcal{M}^2 dt \right] \hat{\rho} \, dt
		+ \mathcal{M} \frac{\sqrt{k}}{2} 
			\left \{  \hat{A}_1 - \langle \hat{A}_1 \rangle ,  \hat{\rho}(t) \right \} dt
	\label{eq:MEQFeedForwardFull}
\end{align}			
 We see there are two terms here that encode the interaction between systems $1$ and $2$ generated by our feedforward protocol.  The first is the  
 $\sqrt{k} \langle \hat{A}_1 \rangle \mathcal{M} $ term, which describes the fact that the average value of $\hat{A}_1$ determines the effective force applied on system $2$.  The second influence is through the last term of the equation; as mentioned above, this describes correlations between the backaction noise (conditioning) of system $1$ and the noise in the feedforward force applied to system $2$.  Note that both these interaction terms scale like $\sqrt{k \gamma_{\rm ff}}$, i.e.~they depend on the product of the measurement strength and the feedforward strength.  
 
 The above equation tells us about the unconditional evolution of the system under the measurement and feedforward protocol.  The evolution must necessarily be directional, as there is no way for system 2 to influence system 1.  We would like to show that this equation is in fact in the same form as the directional master equation of Eq.~(\ref{eq:MEQAsymmetric}) with the choices $\hat{O}_1 = \hat{A}_1$, $\hat{O}_2 = \hat{F}_2$. With these identifications, we expect a master equation:
\begin{align}
	\frac{d}{dt} \hat{\rho} & = 
		-i \lambda  [ \hat{A}_1 \hat{F}_2   , \hat{\rho} ]
		+ \lambda \mathcal{D}[ \eta^{1/2} \hat{A}_1 -  i  \eta^{-1/2}  \hat{F}_2 ] \hat{\rho} \\
		& = 
				-i \lambda  [ \hat{A}_1 \hat{F}_2   , \hat{\rho} ]
		+ \lambda \eta \mathcal{D}[\hat{A}_1] +  \frac{\lambda}{ \eta}  \mathcal{D}[\hat{F}_2] 
			- i \lambda \left( \hat{F}_2 \hat{\rho} \hat{A}_1 - \hat{A}_1 \hat{\rho} \hat{F}_2 \right) 
			\label{eq:TargetMEQ}
\end{align}
In the last line, we have made use of the fact that both $\hat{A}_1$ and $\hat{F}_2$ are Hermitian.  As we already remarked earlier, the Hermiticity of both these operators implies that the ``dissipative interaction" between the two systems does not correspond to the non-Hermitian Hamiltonian terms in our master equation, but rather to jump terms (where there are operators acting on both sides of $\hat{\rho}$).

First, consider the term in Eq.~(\ref{eq:MEQFeedForwardFull}) that is quadratic in the operator $\hat{F}_2$.  This is the $\mathcal{M}^2 dt$ term above, a term describing evolution due to noise in the feedforward propagator.  A straightforward evaluation shows:
\begin{equation}
	\frac{1}{2} \mathcal{M}^2 \hat{\rho} = \gamma_{\rm ff} \mathcal{D}[\hat{F}_2] \hat{\rho}
\end{equation}
This indeed corresponds to the $\hat{F}_2^2$ term in Eq.~(\ref{eq:TargetMEQ}), if we set:
\begin{equation}
	\lambda / \eta =\gamma_{\rm ff}
\end{equation}

Next, consider the most interesting terms in Eq.~(\ref{eq:MEQFeedForwardFull}), those that are linear in $\hat{F}_2$.  These describe the effective interaction that results from the feedforward protocol, where system 2 evolves in a way that is determined by the results of the system 1 measurement.  The order $\hat{F}_2$ terms are:
\begin{align}
	\frac{\sqrt{k}}{2} \mathcal{M} 
		\left( 
			\hat{A}_1 \hat{\rho} + \hat{\rho} \hat{A}_1 
		\right) 
	& = 
		\frac{\sqrt{k}}{2} (-i) \sqrt{\gamma_{\rm ff}} \left[ 
			\hat{F}_2, \hat{A}_1 \hat{\rho}+ + \hat{\rho} \hat{A}_1
				\right] \\
	& = 
		-i \frac{\sqrt{k \gamma_{\rm ff} }}{2} \left( \hat{F}_2 \hat{\rho} \hat{A}_1 - \hat{A}_1  \hat{\rho} \hat{F}_2 \right) 
		-i \frac{\sqrt{k \gamma_{\rm ff}}}{2} [ \hat{A}_{1} \hat{F}_2, \hat{\rho} ] 
\end{align}
These are in complete agreement with the order $\hat{F}_2$ terms in Eq.~(\ref{eq:TargetMEQ}), which describe both coherent and dissipative interactions; we only need to make the identification:
\begin{equation}
	\lambda = \frac{\sqrt{k \gamma_{\rm ff}}}{2} 
\end{equation}

Finally, it is easy to confirm that the remaining $\hat{A}_1^2$ terms are also in agreement with Eq.~(\ref{eq:TargetMEQ}) with the same assignment of parameters.  Summarizing, we see that our general non-reciprocal master equation in Eq.~(\ref{eq:MEQAsymmetric}) describes quantum measurement plus feedforward with the identifications:
\begin{equation}
	\lambda = \sqrt{  \frac{k \gamma_{\rm ff}}{4} }, \,\,\,\,\,  \eta = \sqrt{\frac{k}{ 4 \gamma_{\rm ff} }}
\end{equation} 
 We thus have two crucial conclusions to make:
\begin{itemize}
	\item  Continuous measurement plus feedforward give unconditional evolution that is equivalent to our general non-reciprocal master equation, with system operators that are purely Hermitian
	\item  A corollary is that if we have a non-reciprocal master equation described by Eq.~(\ref{eq:MEQAsymmetric}) with Hermitian operators, then it can {\it always} be interpreted in terms of a measurement plus feedforward protocol
\end{itemize}
In the case where both system operators in the master equation are Hermitian, we see that the interaction strength $\lambda$ is the geometric mean of the measurement and feedforward strengths, whereas the asymmetry parameter is set by their ratio.  

\subsection{Entanglement generation via non-reciprocal interactions}

A basic question about quantum non-reciprocal interactions is whether they are able to generate entanglement between two systems.  Consider the very general master equation in Eq.~(\ref{eq:MEQAsymmetric}) that describes a fully non-reciprocal interaction between two systems $1$ and $2$.  We could add to this dynamics purely local Hamiltonians for system 1 and 2, and then ask whether this equation can generate entanglement.  Specifically, if the system starts in a state that is a product state, can this dynamics ever create a truly entangled state at later times (i.e.~a density matrix that cannot be written as a statistical mixture of product states)?  We know that in general interactions between two systems can generate entanglement.  However, given that dissipation is a crucial part of our non-reciprocal interactions, one might worry that the associated quantum noise might disrupt entanglement generation.  

The above connection to measurement plus feedforward lets us say something crucial about the above question:  if the operators $\hat{O}_1$ and $\hat{O}_2$ that define the non-reciprocal interaction are Hermitian, then \emph{there can never be any generation of entanglement between systems $1$ and $2$}.  In this case, the non-reciprocal interaction is completely equivalent to the interaction generated by a measurement plus feedforward protocol.  Such a protocol only involves local operations and classical communication between the two systems (i.e. LOCC), hence entanglement generation is impossible.  

The corollary is that if both coupling operators $\hat{O}_1$ and $\hat{O}_2$ are non-Hermitian, then there is no general mapping to an LOCC measurement plus feedforward protocol, and hence it is possible to generate entanglement.  In this case, one can always think of the non-reciprocal interaction using the cascaded quantum systems picture, where we have an auxiliary unidirectional waveguide which mediates the interactions between systems 1 and 2.  In general, it is indeed possible for such an interaction to transport particles from system $1$ to $2$, resulting in the generation of entanglement, and even the stabilization of pure entangled states.  For a concrete example, consider the case where both systems $1$ and $2$ are bosonic cavities, and we take $\hat{O}_1 = \hat{a}_1, \hat{O}_2^\dagger = \hat{a}_2$ (with $\hat{a}_j$ the photon annihilation operator for cavity $j$).  Eq.~(\ref{eq:MEQAsymmetric}) then describes a standard cascaded quantum systems coupling, where photons can leak out of cavity $2$, enter the chiral waveguide, and then reach cavity $1$ (but not the reverse process).  By adding local drives to the two cavities (e.g. paramteric two photon drives), such a master equation can indeed generate entanglement, and even stabilize pure entangled states \cite{Mamaev2018}.  Similar constructions are possible with qubits \cite{Stannigel2012}.

Hence, we see that in general, non-reciprocal interactions described by the master equation Eq.~(\ref{eq:MEQAsymmetric}) are  not equivalent to an LOCC measurement plus feedforward protocol.  We see that despite the dissipative nature of these interactions, they are indeed capable of generating quantum entanglement between systems 1 and 2.  Note that the connections between correlated dissipative processes and entanglement generation was recently studied in Ref.~\cite{Seif2021}, in the context of correlated Markovian dephasing in a many qubit system.  

Finally, one might be confused by the above statements, as it seems like we could always connect the general case of non-Hermitian coupling operators to the Hermitian case.  For concreteness, we could always write $\hat{O}_j = \hat{X}_j + i \hat{P}_j$, where both $\hat{X}_j$, $\hat{P}_j$ are Hermitian.  In this case, the coherent interaction Hamiltonian in our master equation becomes:
\begin{align}
	\hat{H}_{\rm coh} & = 
		\frac{\lambda}{2} \left( \hat{O}_1 \hat{O}_2 + \textrm{h.c.} \right) 
		 = 
			\lambda \left( \hat{X}_1 \hat{X}_2 - \hat{P}_1 \hat{P}_2 \right)
\end{align}
This looks like we have two interaction channels now between systems 1 and 2, each described by a pair of Hermitian operators.  One might erroneously conclude that the non-reciprocal interaction described by Eq.~(\ref{eq:MEQAsymmetric}) could always be written in terms of a {\it pair} of non-reciprocal interactions, each corresponding to Hermitian coupling operators.  

This is incorrect.  While $\hat{H}_{\rm coh}$ can indeed be decomposed like this, the same is not true for the dissipative interactions encoded by the correlated dissipator in Eq.~(\ref{eq:MEQAsymmetric}).  To be clear, in general
\begin{equation}
	\mathcal{D}[\hat{O}_1 \mp i \hat{O}_2^\dagger ] 
		\neq
	\mathcal{D}[\hat{X}_1 \mp i \hat{X}_2^\dagger ] +
	\mathcal{D}[\hat{P}_1 \mp i \hat{P}_2^\dagger ] 
\end{equation}
Hence, non-reciprocal quantum interactions realized using non-Hermitian coupling operators are fundamentally different from the case with Hermitian operators (which can always be reduced to a LOCC measurement plus feedforward scheme).

\section{Conclusion}

These lectures have attempted to provide an intuitive picture for how external driving, dissipation and interference can be harnessed to realize non-reciprocal interactions and devices in a fully consistent quantum mechanical setting.  Using a simple toy model, we showed explicit connections between seemingly different formalisms:  Hamiltonians with synthetic gauge fields, non-Hermitian effective Hamiltonians, and quantum master equations.  We also discussed the explicit connection between this approach to non-reciprocity and th effective evolution generated by quantum measurement and feedforward schemes.  We hope the discussion here will assist researchers both in designing new kinds of non-reciprocal devices, as well as investigations of the fundamental quantum many-body physics of systems whose underlying interactions are directional.  

\section*{Acknowledgements}

I am grateful to the many people over the years that have helped me better understand quantum non-reciprocity and have helped develop ideas on this topic:  Michel Devoret, Steve Girvin, Florian Marquardt, Alexander McDonald, Oskar Painter, Chen Wang and Yuxin Wang.  I am of course especially grateful to Anja Metelemann, with whom the main ideas discussed in these notes were developed.  I also want to thank Yuxin Wang for her assistance in helping prepare figures and for carefully proofreading and critiquing this set of notes.  This work was supported by Army Research Office under grant W911-NF-19-1-0380, and by the Air Force Office of Scientific Research MURI program under grant no. FA9550-19-1-0399.


\paragraph{Funding information}
Authors are required to provide funding information, including relevant agencies and grant numbers with linked author's initials. Correctly-provided data will be linked to funders listed in the \href{https://www.crossref.org/services/funder-registry/}{\sf Fundref registry}.

\begin{appendix}

\section{Inteference cancellation to all orders in $t$}
\label{app:HigherOrderInterference}

A crucial but surprising conclusion in the main text was that the simple perturbative interference condition in Eq.~(\ref{eq:NRCondPhi}) and (\ref{eq:NRCondKappa}) that causes $G^R[2,1;\omega]$ to vanish by destructive interfernce is in fact valid to all orders in $t$.  We formalize the argument here provided in the main text.   First, defining $\tilde{\omega} = \omega + i \kappa/2$, we have:
\begin{align}
	G^R[2,1;\omega] = \left[ \left(\tilde{\omega} - \bm{H} \right)^{-1} \right]_{21} 
		= \frac{1}{\tilde{\omega}} \left[ \sum_{m=0}^\infty \left( \frac{\bm{H}}{\tilde{\omega}} \right)^m \right]_{11}
\end{align}
where $H$ is the $3 \times 3$ Hermitian Hamiltonian matrix that encodes hopping (c.f.~Eq.~(\ref{eq:NonHermMatrix})).  
Expanding out the product here, we can write:
\begin{align}
	G^R[2,1;\omega] & = 
		 \frac{1}{\tilde{\omega}}  \left[  \frac{\bm{H}}{\tilde{\omega}} \right]_{21}
			 \left[ \sum_{m=0}^\infty \left( \frac{\bm{H}}{\tilde{\omega}} \right)^m \right]_{11}
			 +
		 \frac{1}{\tilde{\omega}}  \left[  \frac{\bm{H}}{\tilde{\omega}} \right]_{23}
			 \left[ \sum_{m=0}^\infty \left( \frac{\bm{H}}{\tilde{\omega}} \right)^m \right]_{31} \\
	&=		 \frac{1}{\tilde{\omega}} \Bigg(
				 \left[  \frac{\bm{H}}{\tilde{\omega}} \right]_{21}
			 \left[ \sum_{m=0}^\infty \left( \frac{\bm{H}}{\tilde{\omega}} \right)^m \right]_{11}
			 + \nonumber \\
&
		  \left[  \frac{\bm{H}}{\tilde{\omega}} \right]_{23}
		 \left[  \frac{\bm{H}}{\tilde{\omega}} \right]_{31}
			 \left[ \sum_{m=0}^\infty \left( \frac{\bm{H}}{\tilde{\omega}} \right)^m \right]_{11} \
			 +
		  \left[  \frac{\bm{H}}{\tilde{\omega}} \right]_{23}
		 \left[  \frac{\bm{H}}{\tilde{\omega}} \right]_{32}
			 \left[ \sum_{m=0}^\infty \left( \frac{\bm{H}}{\tilde{\omega}} \right)^m \right]_{21}  
			 \Bigg)
\end{align}
Now, re-express the geometric series in each term in terms of the retarded Green's function:
\begin{align}
	G^R[2,1;\omega] & = 
	 \left[  \frac{\bm{H}}{\tilde{\omega}} \right]_{21} G^R[1,1;\omega] +
	  \left[  \frac{\bm{H}}{\tilde{\omega}} \right]_{23}
		 \left[  \frac{\bm{H}}{\tilde{\omega}} \right]_{31} G^R[1,1;\omega] +
	  \left[  \frac{\bm{H}}{\tilde{\omega}} \right]_{23}
		 \left[  \frac{\bm{H}}{\tilde{\omega}} \right]_{32} G^R[2,1;\omega]  
\end{align}		 

Solving this equation for $G^R[2,1,\omega]$ yields:
\begin{align}		
		 G^R[2,1;\omega] & =  Z[2,2;\omega] \left( \left[  \frac{\bm{H}}{\tilde{\omega}} \right]_{21}  +
	  \left[  \frac{\bm{H}}{\tilde{\omega}} \right]_{23}
		 \left[  \frac{\bm{H}}{\tilde{\omega}} \right]_{31}  \right)  G^R[1,1;\omega] 
	\label{eq:FullGRInterference}
\end{align}
with
\begin{align}
			Z[2,2;\omega] = \left( 1 -    \left[  \frac{\bm{H}}{\tilde{\omega}} \right]_{23}
		 \left[  \frac{\bm{H}}{\tilde{\omega}} \right]_{32} \right)^{-1}
\end{align}
The two terms on the RHS in Eq.~(\ref{eq:FullGRInterference}) exactly correspond to the amplitudes $Q_{1,tot}$ and $Q_{2,tot}$ in Eqs.~(\ref{eq:Q1tot}) and (\ref{eq:Q2tot}).  We thus see that the even including processes to all orders in the tunneling, the simple interference condition of Eqs.~(\ref{eq:NRCondPhi}) and (\ref{eq:NRCondKappa}) guarantees that $G^R[2,1;\omega] = 0$.

\end{appendix}

%




\bibliography{LesHouches2019ACBib.bib}

\nolinenumbers

\end{document}